\documentclass[12pt]{article}
\pdfoutput=1
\usepackage{latexsym}
\usepackage{amssymb, amsfonts, amsmath}
\usepackage{graphicx} 
\usepackage{indentfirst}
\usepackage{bbm}
\usepackage{amssymb}
\usepackage{verbatim}
\usepackage{amsmath, amsthm, amssymb}
\usepackage{mathrsfs}
\usepackage{amsfonts}
\usepackage{dsfont}
\usepackage{csquotes}
\usepackage{hyperref}
\usepackage{url}
\usepackage{xcolor}
\usepackage{xpatch}
\usepackage[english]{babel}

\usepackage[style=phys, articletitle=true, biblabel=brackets, backend=bibtex,
chaptertitle=false, pageranges=false, eprint=true,%
natbib=true, maxbibnames=10, giveninits=true, sorting=none]{biblatex} 

\DeclareFieldFormat[article]{journaltitle}{\mkbibitalic{#1\isdot}} 
\makeatletter
\newrobustcmd{\mkbibfixedbrackets}[1]{%
	\begingroup
	\blx@blxinit
	\blx@setsfcodes
	\bibleftbracket#1\bibrightbracket
	\endgroup}

\renewbibmacro*{doi+eprint+url}{%
	\iftoggle{bbx:doi}
	{\printfield{doi}}
	{}%
	\ifboolexpr{test {\iffieldequalstr{eprinttype}{arxiv}} or test {\iffieldequalstr{eprinttype}{arXiv}}}
	{\setunit{\addspace}\newblock}
	{\newunit\newblock}%
	\iftoggle{bbx:eprint}
	{\usebibmacro{eprint}}
	{}%
	\newunit\newblock
	\iftoggle{bbx:url}
	{\usebibmacro{url+urldate}}
	{}}

\xpatchbibdriver{online}
{\newunit\newblock
	\iftoggle{bbx:eprint}
	{\usebibmacro{eprint}}
	{}}
{\ifboolexpr{test {\iffieldequalstr{eprinttype}{arxiv}} or test {\iffieldequalstr{eprinttype}{arXiv}}}
	{\setunit{\addspace}\newblock}
	{\newunit\newblock}%
	\iftoggle{bbx:eprint}
	{\usebibmacro{eprint}}
	{}}
{}{}

\DeclareFieldFormat{eprint:arxiv}{%
	\mkbibbrackets{%
		\ifhyperref
		{\href{http://arxiv.org/\abx@arxivpath/#1}{%
				arXiv\addcolon
				\nolinkurl{#1}%
				\iffieldundef{eprintclass}
				{}
				{\addspace\UrlFont{\mkbibfixedbrackets{\thefield{eprintclass}}}}}}
		{arXiv\addcolon
			\nolinkurl{#1}%
			\iffieldundef{eprintclass}
			{}
			{\addspace\UrlFont{\mkbibfixedbrackets{\thefield{eprintclass}}}}}}}
\makeatother

\parskip=\medskipamount

\newcommand{\cA}{{\cal A}}

\newcommand{\cF}{{\cal F}}

\newcommand{\cH}{{\cal H}}
\newcommand{\cI}{{\cal I}}

\newcommand{\cN}{{\cal N}}

\newcommand{\cT}{{\cal T}}

\newcommand{\cZ}{{\cal Z}}

\def\a{\alpha}

\def\b{\beta}

\def\d{\delta}
\def\e{\epsilon}
\def\f{\phi}
\def\g{\gamma}
\def\G{\Gamma}

\def\l{\lambda}

\def\s{\sigma}

\def\D{\Delta}
\def\F{\Phi}
\def\J{\Psi}

\def\P{\Pi}

\newcommand{\ad}{{\dot{\alpha}}}                           
\newcommand{\bd}{{\dot{\beta}}}                            
\newcommand{\ve}{\varepsilon}                            

\newcommand{\pa}{\partial}                           

\newcommand{\be}{\begin{equation}}
\newcommand{\ee}{\end{equation}}
\newcommand{\bea}{\begin{eqnarray}}
\newcommand{\eea}{\end{eqnarray}}

\newcommand{\ba}{\begin{array}}
\newcommand{\ea}{\end{array}}

\def\double #1{#1{\hbox{\kern-2pt $#1$}}}

\newcommand{\gd}{{\dot\g}}

\newcommand{\bsubeq}{\begin{subequations}}
\newcommand{\esubeq}{\end{subequations}}

\numberwithin{equation}{section}

\begin{document}

\begin{titlepage}
\begin{flushright}
Apr, 2022
\end{flushright}
\vspace{2mm}

\begin{center}
\Large \bf Three-point functions of a fermionic higher-spin \\ current in 4D conformal field theory
\end{center}

\begin{center}
{\bf
Evgeny I. Buchbinder and Benjamin J. Stone}

{\footnotesize{
{\it Department of Physics M013, The University of Western Australia\\
35 Stirling Highway, Crawley W.A. 6009, Australia}} ~\\
}
\end{center}
\begin{center}
\texttt{Email: evgeny.buchbinder@uwa.edu.au, \\ benjamin.stone@research.uwa.edu.au}
\end{center}

\vspace{4mm}

\begin{abstract}
\baselineskip=14pt


We investigate the properties of a four-dimensional conformal field theory possessing a fermionic higher-spin 
current $Q_{\a(2k) \ad}$. 
Using a computational approach, we examine the number of independent tensor structures contained in the three-point correlation functions of two fermionic higher-spin currents with the conserved vector current $V_m$, and with the energy-momentum tensor $T_{mn}$. In particular, the $k=1$ case corresponds to a ``supersymmetry-like" current, that is, a 
fermionic conserved current with identical properties to the supersymmetry current which appears
in $\cN=1$ superconformal field theories. However, we show that in general, the three-point correlation functions $\langle Q Q T\rangle $, $\langle \bar{Q} Q V\rangle $ and $\langle \bar{Q} Q T\rangle $
are not consistent with $\cN=1$ supersymmetry. 


\end{abstract}
\end{titlepage}

\newpage
\renewcommand{\thefootnote}{\arabic{footnote}}
\setcounter{footnote}{0}

\tableofcontents
\vspace{1cm}
\bigskip\hrule


\section{Introduction}\label{section1}


Correlation functions of conserved currents are among the most important observables in conformal field theory. 
It is a well known fact that conformal symmetry determines the general form of two- and three-point correlation functions up to finitely many parameters, however, it remains an open problem to understand the structure of three-point functions of conserved currents for arbitrary spin. The systematic approach to study correlation functions of conserved currents was undertaken in~\cite{Osborn:1993cr, Erdmenger:1996yc} 
(see also refs.~\cite{Polyakov:1970xd, Schreier:1971um, Migdal:1971xh, Migdal:1971fof, Ferrara:1972cq, 
Ferrara:1973yt, Koller:1974ut, Mack:1976pa, Stanev:1988ft, Fradkin:1978pp} for earlier results), and
was later extended to superconformal field theories in diverse dimensions \cite{Park:1997bq, Osborn:1998qu, Park:1999pd, Park:1998nra, Park:1999cw, Kuzenko:1999pi, 
Nizami:2013tpa, Buchbinder:2015qsa, Buchbinder:2015wia, Kuzenko:2016cmf, Buchbinder:2021gwu, Buchbinder:2021izb, Buchbinder:2021kjk, Buchbinder:2021qlb}.\footnote{The approach of~\cite{Osborn:1993cr, Erdmenger:1996yc} 
performs the analysis in general dimensions and did not consider parity-violating structures relevant for three-dimensional conformal field theories. 
These structures were found later in~\cite{Giombi:2011rz}.} The most important examples of conserved currents in conformal field theory are the energy-momentum tensor and vector currents; their three-point functions were studied in \cite{Osborn:1993cr}. However, more general conformal field theories can possess higher-spin conserved currents. As was proven by Maldacena and Zhiboedov in~\cite{Maldacena:2011jn},
all correlation functions of higher-spin currents are equal to those of a free theory. This theorem was originally proven in three dimensions 
and was later generalised in~\cite{Stanev:2013qra, Alba:2013yda, Alba:2015upa} to four- and higher-dimensional cases. The general structure 
of the three-point functions of conserved higher-spin, bosonic, vector currents was found by Stanev~\cite{Stanev:2012nq} and 
Zhiboedov~\cite{Zhiboedov:2012bm}, see also \cite{Elkhidir:2014woa} for similar results in the embedding formalism~\cite{Weinberg:2010fx, Weinberg:2012mz, Costa:2011dw, Costa:2011mg, Costa:2014rya, Fortin:2020des} (and \cite{Goldberger:2011yp, Goldberger:2012xb} for supersymmetric extensions). There are also some novel approaches to the construction of correlation functions of conserved currents which carry out the calculations in momentum space, using methods such as spinor-helicity variables \cite{Jain:2020puw, Jain:2020rmw, Jain:2021gwa, Jain:2021vrv, Jain:2021wyn, Isono:2019ihz, Bautista:2019qxj}.

The study of correlations functions in conformal field theory has mostly been devoted to bosonic operators with vector indices (except for supersymmetric settings);
fermionic operators have practically not been studied.\footnote{Recently, in~\cite{Elkhidir:2014woa}, correlation functions involving fermionic operators were studied, however	these operators were not conserved currents.} Our interest in studying three-point functions of fermionic operators is two-fold; first, any conformal field theory possessing fermionic operators
naturally breaks the assumptions of the Maldacena--Zhiboedov theorem~\cite{Maldacena:2011jn} discussed above. Indeed, the main assumption of the 
Maldacena--Zhiboedov theorem was that the conformal field theory under consideration possesses a unique conserved current of spin two, the energy-momentum tensor. However, in~\cite{Maldacena:2011jn} it was also shown that if a conformal field theory possesses a conserved fermionic higher-spin current then it has an additional conserved current of spin two. 
Hence, it is not clear whether correlation functions of fermionic higher-spin currents must coincide with those in a free theory. Second, fermionic operators are interesting due to their prevalence in supersymmetric field theories. 
In fact, there is a natural question: if a conformal field theory possesses a conserved fermionic current, is it necessarily supersymmetric?

The aim of this paper is to study correlation functions of the conserved fermionic higher-spin currents\footnote{We use the standard notation $\F_{\a(m)  \ad(n)} = \F_{(\a_{1} \dots \a_{m}) (\ad_{1} \dots \ad_{n}) }$.}
\begin{equation}
	Q_{\a(2 k)  \ad}\,, \hspace{10mm} \bar{Q}_{\a \ad (2 k)} \, , \label{0.1}
\end{equation}
which obey the conservation equations
\begin{equation}
	\partial^{\a \ad} Q_{\a \a (2k-1) \ad} = 0 \, , \hspace{10mm} \partial^{\a \ad} \bar{Q}_{\a \ad \ad(2k-1)} = 0 \, .	\label{0.11}
\end{equation}
The case $k=1$ in~\eqref{0.1} is quite interesting as it corresponds to currents of spin--$\frac{3}{2}$ which possess the same index structure and conservation properties as the supersymmetry currents. Indeed, one might expect that a conformal field theory possessing conserved spin--$\frac{3}{2}$ primary operators is supersymmetric. One way to explore this issue is to study the correlation functions involving such operators to see if they are consistent with supersymmetry. In particular, we must study the general form of the three-point functions involving combinations of the operators $Q_{\a(2) \ad}, \bar{Q}_{\a \ad(2)} $ (i.e. \eqref{0.1} for $k=1$), the energy-momentum tensor $T_{m n}$ and the vector current $V_{m}$. Recall that in any superconformal field theory the supersymmetry current and the energy-momentum tensor are components of the supercurrent multiplet, $J_{\a \ad}(z)$, where $z = (x^{m}, \theta^{\a}, \bar{\theta}_{\ad} )$ is a point in 4D Minkowski superspace. This implies that in supersymmetric theories the three-point functions 
\begin{equation} 
	\langle \bar{Q}_{\a \ad (2)} (x_1) \,  Q_{\b(2)  \dot{\b}} (x_2) \, T_{mn} (x_3) \rangle\,, \hspace{10mm}
	\langle Q_{\a (2 ) \ad} (x_1) \,  Q_{\b(2 )  \bd} (x_2) \, T_{mn} (x_3) \rangle \, , 
	\label{0.2}
\end{equation}
must be contained in the three-point function of the supercurrent $ \langle J_{\a \ad} (z_1) \, J_{\b \bd} (z_2) \, J_{\g \gd} (z_3) \rangle $, which was shown in \cite{Osborn:1998qu} to be fixed up to two independent tensor structures. 
Similarly, in supersymmetric theories the vector current $V_{m}$ is a component of the flavour current multiplet, $L(z)$. Hence, the three-point functions 
\begin{equation} 
	\langle \bar{Q}_{\a \ad(2 )} (x_1) \, Q_{\b(2 )  \bd} (x_2) \, V_{m} (x_3) \rangle\,, \hspace{10mm}
	\langle Q_{\a (2) \ad} (x_1) \, Q_{\b(2 )  \bd} (x_2) \, V_{m} (x_3) \rangle \, , 
	\label{0.3}
\end{equation}
must be contained in the three-point function of the supercurrent and the flavour current $\langle J_{\a \ad} (z_1) \, J_{\b \bd} (z_2) \, L (z_3) \rangle $, which was shown to be fixed up to a single tensor structure \cite{Osborn:1998qu}.

In this paper, we study the general form of the three-point functions \eqref{0.2}, \eqref{0.3} and extend the results to the operators \eqref{0.1}, using only the constraints of conformal symmetry; supersymmetry is not assumed. The analysis is highly non-trivial and requires significant use of computational methods. To streamline the calculations we develop a hybrid formalism which combines the approach of Osborn and Petkou~\cite{Osborn:1993cr} and the approach based on contraction of tensor indices with auxiliary vectors/spinors. This method is widely used throughout the literature to construct correlation functions of more complicated tensor operators. Our particular approach, however, has some advantages as the correlation function is completely described in terms of a polynomial which is a function of a single conformally covariant three-point building block, $X$, and the auxiliary spinor variables $u, \bar{u}, v, \bar{v}, w, \bar{w}$. Hence, one does not have to work with the spacetime points explicitly when imposing conservation equations. To find all solutions for the polynomial, we construct a generating function which produces an exhaustive list of all possible linearly dependent structures for fixed (and in some cases, arbitrary) spins. The possible structures form a basis in which the polynomial may be decomposed, and are in correspondence with the solutions to a set of six linear inhomogeneous Diophantine equations, which can be solved computationally for any spin. 

Using the methods outlined above, we find that the three-point functions~\eqref{0.2}, \eqref{0.3}, 
in general, are not consistent with supersymmetry as they are fixed up to more independent tensor structures than the three-point functions $\langle J  J  J \rangle $ and  
$\langle J J L \rangle $. This means, based on the constraints of conformal symmetry alone, that the existence of spin--$\frac{3}{2}$ ``supersymmetry-like" conserved currents 
in a conformal field theory does not necessarily imply that the theory is superconformal. We want to stress that our analysis is based only on symmetries and does not 
take into account other features of local field theory. We do not know how to realise a local non-supersymmetric conformal field theory possessing conserved spin--$\frac{3}{2}$ currents, neither do we have a proof that it is impossible. 

Our paper is organised as follows: in Section \ref{section2}, we discuss the general formalism to construct two- and three-point functions in
conformal field theory. First, we review the constructions of Osborn and Petkou~\cite{Osborn:1993cr} and introduce our hybrid generating function formalism based on contractions of tensor operators with auxiliary spinors. We construct a generating function which, for a given choice of spins, generates all possible linearly dependent solutions for the correlation function. In Sections \ref{section3} and \ref{section4}, we find the most general form of the three-point functions~\eqref{0.3}. Our conclusions are that the three-point function $ \langle \bar{Q} Q V \rangle$ depends on three independent
tensor structures (here and in all other cases the structures are found explicitly), while the three-point function $ \langle Q Q V \rangle$ 
vanishes in general. In Sections \ref{section5} and \ref{section6}, we find the most general form 
of the three-point functions~\eqref{0.2}. Our conclusions are that the three-point function $ \langle \bar{Q} Q T \rangle$ is determined up to four independent
tensor structures and the three-point function $ \langle Q Q T \rangle$ is fixed up to a single tensor structure. Most of our analysis in Sections \ref{section3}--\ref{section6}
was performed for an arbitrary $k$. However, due to computational limitations certain results were proven only for small values $k$. Nevertheless, we
believe that the results stated above hold for all values of $k$. Finally, in Section \ref{Discussion}, we discuss whether our results are consistent with supersymmetry for $k=1$, 
when $Q$ possesses the same properties as the supersymmetry current. We show that, in general, the results obtained in Sections \ref{section3}--\ref{section6} are not consistent with supersymmetry. Our four-dimensional notation and conventions are summarised in Appendix \ref{AppA}.




\section{Conformal building blocks}\label{section2}

In this section we will review the pertinent aspects of the group theoretic formalism used to compute correlation functions of primary operators in four dimensional conformal field theories. For a more detailed review of the formalism as applied to correlation functions of bosonic primary fields, the reader may consult \cite{Osborn:1993cr}. Our 4D conventions and notation are those of \cite{Buchbinder:1998qv}, see the Appendix \ref{AppA} for a brief overview.

\subsection{Two-point functions}\label{subsection2.2}

Consider 4D Minkowski space $\mathbb{M}^{1,3}$, parameterised by coordinates $ x^{m} $, where $m = 0, 1, 2, 3$ are Lorentz indices. Given two points, $x_{1}$ and $x_{2}$, we can define the covariant two-point function
\begin{equation} \label{Two-point building blocks 1}
	x_{12}^{m} = (x_{1} - x_{2})^{m} \, , \hspace{10mm} x_{21}^{m} = - x_{12}^{m} \, . 
\end{equation}
Next, following Osborn and Petkou \cite{Osborn:1993cr}, we introduce the conformal inversion tensor, $I_{mn}$, which is defined as follows:
\begin{align} \label{Inversion tensor}
I_{mn}(x) = \eta_{mn} - 2 \, \frac{ x_{m} x_{n}}{x^{2}} \, , \hspace{10mm} I_{m a}(x) \, I^{a n}(x) = \d_{m}^{n} \, .
\end{align}
This object played a pivotal role in the construction of correlation functions in \cite{Osborn:1993cr}, as the full conformal group may be generated by 
considering Poincar\'e transformations supplemented by inversions. However, in the context of this work, we require an analogous operator for the spinor representation. 
Hence, we convert the vector two-point functions \eqref{Two-point building blocks 1} into spinor notation using the conventions outlined in appendix \ref{AppA}:
\begin{align}
	x_{12 \, \a \ad} &= (\s^{m})_{\a \ad} x_{12 \, m} \, , & x_{12}^{\ad \a} &= (\tilde{\s}^{m})^{\ad \a} x_{12 \, m} \, , & x_{12}^{2} &= - \frac{1}{2} x_{12}^{\ad \a} x_{12 \, \a \ad} \, .
\end{align}
In this form the two-point functions possess the following useful properties:
\begin{align}  \label{Two-point building blocks - properties 1} 
 	x_{12}^{\ad \a} x_{12 \, \b \ad} = - x_{12}^{2} \d_{\b}^{\a} \, , \hspace{10mm} x_{12}^{\ad \a} x_{12 \, \a \bd} = - x_{12}^{2} \d_{\bd}^{\ad} \, . 
\end{align}
Hence, we find
\begin{equation} \label{Two-point building blocks 4}
	(x_{12}^{-1})^{\ad \a} = - \frac{x_{12}^{\ad \a}}{x_{12}^{2}} \, .
\end{equation}
We also introduce the normalised two-point functions, denoted by $\hat{x}_{12}$,
\begin{align} \label{Two-point building blocks 3}
		\hat{x}_{12 \, \a \ad} = \frac{x_{12 \, \a \ad}}{( x_{12}^{2})^{1/2}} \, , \hspace{10mm} \hat{x}_{12}^{\ad \a} \hat{x}_{12 \, \b \ad} = - \d_{\a}^{\b} \, . 
\end{align}
From here we can now construct an operator analogous to the conformal inversion tensor acting on the space of symmetric traceless tensors of arbitrary rank. Given a two-point function $x$, we define the operator
\begin{equation} \label{Higher-spin inversion operators a}
\cI_{\a(k) \ad(k)}(x) = \hat{x}_{(\a_{1} (\ad_{1}} \dots \hat{x}_{ \a_{k}) \ad_{k})}  \, ,
\end{equation}
along with its inverse
\begin{equation} \label{Higher-spin inversion operators b}
\bar{\cI}^{\ad(k) \a(k)}(x) = \hat{x}^{(\ad_{1} (\a_{1}} \dots \hat{x}^{ \ad_{k}) \a_{k})} \, .
\end{equation}
The spinor indices may be raised and lowered using the standard conventions as follows:
\begin{subequations}
	\begin{align}
		\cI^{\a(k)}{}_{\ad(k)}(x) &= \ve^{\a_{1} \g_{1}} \dots \ve^{\a_{k} \g_{k}} \, \cI_{\g(k) \ad(k)}(x) \, , \\
		\bar{\cI}_{\ad(k)}{}^{\a(k)}(x) &= \ve_{\ad_{1} \gd_{1}} \dots \ve_{\ad_{k} \gd_{k}} \, \bar{\cI}^{\gd(k) \a(k)}(x) \, .
	\end{align}
\end{subequations}
Now due to the property
\begin{equation}
	\cI_{\a(k) \ad(k)}(-x) = (-1)^{k} \cI_{\a(k) \ad(k)}(x) \, ,
\end{equation}
we have the following useful relations:
\begin{subequations} \label{Higher-spin inversion operators - properties}
	\begin{align}
		\cI_{\a(k) \ad(k)}(x_{12}) \, \bar{\cI}^{\ad(k) \b(k)}(x_{21}) &= \d_{(\a_{1}}^{(\b_{1}} \dots \d_{\a_{k})}^{\b_{k})} \, , \\
		\bar{\cI}^{\bd(k) \a(k)}(x_{12}) \, \cI_{\a(k) \ad(k)}(x_{21}) &= \d_{(\ad_{1}}^{(\bd_{1}} \dots \d_{\ad_{k})}^{\bd_{k})} \, .
	\end{align}
\end{subequations}
The objects \eqref{Higher-spin inversion operators a}, \eqref{Higher-spin inversion operators b} prove to be essential in the construction of correlation functions of primary operators with arbitrary spin. Indeed, the vector representation of the inversion tensor may be recovered in terms of the spinor two-point functions as follows:
\begin{equation}
	I_{m n}(x) = - \frac{1}{2} \, \text{Tr}( \tilde{\s}_{m} \, \hat{x} \, \tilde{\s}_{n} \, \hat{x} ) \, .
\end{equation}
%
%
Now let $\F_{\cA}$ be a primary field with dimension $\D$, where $\cA$ denotes a collection of Lorentz spinor indices. The two-point correlation function of $\F_{\cA}$ and its conjugate $\bar{\F}^{\bar{\cA}}$ is fixed by conformal symmetry to the form
\begin{equation}
	\langle \F_{\cA}(x_{1}) \, \bar{\F}^{\bar{\cA}}(x_{2}) \rangle = c \, \frac{\cI_{\cA}{}^{\bar{\cA}}(x_{12})}{(x_{12}^{2})^{\D}} \, , 
\end{equation} 
where $\cI$ is an appropriate representation of the inversion tensor and $c$ is a constant complex parameter. The denominator of the two-point function is determined by the conformal dimension of $\F_{\cA}$, which guarantees that the correlation function transforms with the appropriate weight under scale transformations. For example, in the case of the fermionic current field $Q_{\a(2k) \ad}$, the two-point function is uniquely fixed to the following form:
\begin{equation}
	\langle Q_{\a(2k) \ad}(x_{1}) \, \bar{Q}^{\bd(2k) \b}(x_{2}) \rangle = c \, \frac{ \cI_{\a(2k)}{}^{\bd(2k)}(x_{12}) \, \bar{\cI}_{\ad}{}^{\b} (x_{12})}{(x_{12}^{2})^{\D(Q)}} \, , 
\end{equation} 
where in this case $\D(Q)$ is fixed by conservation of $Q$, $(\bar{Q})$ at $x_{1}$, $(x_{2})$. It is not too difficult to show that $\D(Q) = k + \tfrac{5}{2}$.


\subsection{Three-point functions}\label{subsection2.3}

Given three distinct points in Minkowski space, $x_{i}$, with $i = 1,2,3$, we define conformally covariant three-point functions in terms of the two-point functions as in \cite{Osborn:1993cr}
\begin{align}
	X_{ij} &= \frac{x_{ik}}{x_{ik}^{2}} - \frac{x_{jk}}{x_{jk}^{2}} \, , & X_{ji} &= - X_{ij} \, ,  & X_{ij}^{2} &= \frac{x_{ij}^{2}}{x_{ik}^{2} x_{jk}^{2} } \, , 
\end{align}
where $(i,j,k)$ is a cyclic permutation of $(1,2,3)$. For example we have
\begin{equation}
	X_{12}^{m} = \frac{x_{13}^{m}}{x_{13}^{2}} - \frac{x_{23}^{m}}{x_{23}^{2}} \, , \hspace{10mm} X_{12}^{2} = \frac{x_{12}^{2}}{x_{13}^{2} x_{23}^{2} } \, .
\end{equation}
There are several useful identities involving the two-point and three-point functions along with the conformal inversion tensor, for example we have the useful algebraic relations
\begin{subequations}
	\begin{align} \label{Inversion tensor identities - vector case 1}
		I_{m}{}^{a}(x_{13}) \, I_{a n}(x_{23}) &= I_{m}{}^{a}(x_{12}) \, I_{a n}(X_{13}) \, ,  & I_{m n}(x_{23}) \, X_{12}^{n} &= \frac{x_{12}^{2}}{x_{13}^{2}} \, X_{13 \, m} \, ,
	\end{align} \vspace{-5mm}
	\begin{align} \label{Inversion tensor identities - vector case 2}
		I_{m}{}^{a}(x_{23}) \, I_{a n}(x_{13}) &= I_{m}{}^{a}(x_{21}) \, I_{a n}(X_{32}) \, ,  & I_{m n}(x_{13}) \, X_{12}^{n} &= \frac{x_{12}^{2}}{x_{23}^{2}} \, X_{32 \, m} \, , 
	\end{align}
\end{subequations}
\\
and the differential identities
\begin{align}
	\pa_{(1) \, m} X_{12 \, n} = \frac{1}{x_{13}^{2}} I_{m n}(x_{13}) \, , \hspace{10mm} \pa_{(2) \, m} X_{12 \, n} = - \frac{1}{x_{23}^{2}} I_{m n}(x_{23}) \, . \label{Inversion tensor identities - vector case 3}
\end{align}
The three-point functions also may be represented in spinor notation as follows:
\begin{equation}
	X_{ij , \, \a \ad} = (\s_{m})_{\a \ad} X_{ij}^{m} \, , \hspace{10mm} X_{ij , \, \a \ad} = (x^{-1}_{ik})_{\a \gd} x_{ij}^{\gd \g} (x^{-1}_{jk})_{\g \ad} \, .
\end{equation}
These objects satisfy properties similar to the two-point functions \eqref{Two-point building blocks - properties 1}. Indeed, 
it is convenient to define the normalised three-point functions $\hat{X}_{ij}$, and the inverses $(X_{ij}^{-1})$,
\begin{equation}
	\hat{X}_{ij , \, \a \ad} = \frac{X_{ij , \, \a \ad}}{( X_{ij}^{2})^{1/2}} \, , \hspace{10mm}	(X_{ij}^{-1})^{\ad \a} = - \frac{X_{ij}^{\ad \a}}{X_{ij}^{2}} \, .
\end{equation}  
Now given an arbitrary three-point building block $X$, it is also useful to construct the following higher-spin operator:
\begin{equation}
	\cI_{\a(k) \ad(k)}(X) = \hat{X}_{ (\a_{1} (\ad_{1}} \dots \hat{X}_{\a_{k}) \ad_{k})}  \, ,
\end{equation}
along with its inverse
\begin{equation}
	\bar{\cI}^{\ad(k) \a(k)}(X) = \hat{X}^{(\ad_{1} (\a_{1}} \dots \hat{X}^{ \ad_{k}) \a_{k})} \, .
\end{equation}
These operators have properties similar to the two-point higher-spin inversion operators \eqref{Higher-spin inversion operators a}, \eqref{Higher-spin inversion operators b}. There are also some useful algebraic identities relating the two- and three-point functions at various points, such as
\begin{equation}
	\cI_{\a \ad}(X_{12})  = \cI_{\a \gd}(x_{13}) \, \bar{\cI}^{\gd \g}(x_{12}) \, \cI_{\g \ad}(x_{23}) \, , \hspace{5mm}  \bar{\cI}^{\ad \g}(x_{13}) \, \cI_{\g \gd}(X_{12})  \, \bar{\cI}^{\gd \a}(x_{13}) = \bar{\cI}^{\ad \a}(X_{32})  \, . \label{Inversion tensor identities - spinor case}
\end{equation}
These identities (and cyclic permutations of them) are analogous to \eqref{Inversion tensor identities - vector case 1}, \eqref{Inversion tensor identities - vector case 2}, and also admit higher-spin generalisations, for example
\begin{equation}
	\bar{\cI}^{\ad(k) \g(k)}(x_{13}) \, \cI_{\g(k) \gd(k)}(X_{12}) \, \bar{\cI}^{\gd(k) \a(k)}(x_{13}) = \bar{\cI}^{\ad(k) \a(k)}(X_{32})  \, . \label{Inversion tensor identities - higher spin case}
\end{equation}
In addition, similar to \eqref{Inversion tensor identities - vector case 3}, there are also the following useful identities:
\begin{equation}
	\pa_{(1) \, \a \ad} X_{12}^{ \dot{\s} \s} = - \frac{2}{x_{13}^{2}} \, \cI_{\a}{}^{\dot{\s}}(x_{13}) \,  \bar{\cI}_{\ad}{}^{\s}(x_{13}) \, , \hspace{5mm} \pa_{(2) \, \a \ad} X_{12}^{ \dot{\s} \s} = \frac{2}{x_{23}^{2}} \, \cI_{\a}{}^{\dot{\s}}(x_{23}) \,  \bar{\cI}_{\ad}{}^{\s}(x_{23}) \, . \label{Three-point building blocks - differential identities}
\end{equation}
These identities allow us to account for the fact that correlation functions of primary fields obey differential constraints which can arise due to conservation equations. Indeed, given a tensor field $\cT_{\cA}(X)$, there are the following differential identities which arise as a consequence of \eqref{Three-point building blocks - differential identities}:
\begin{subequations}
	\begin{align}
		\pa_{(1) \, \a \ad} \cT_{\cA}(X_{12}) &= \frac{1}{x_{13}^{2}} \, \cI_{\a}{}^{\dot{\s}}(x_{13}) \,  \bar{\cI}_{\ad}{}^{\s}(x_{13}) \, \frac{ \pa}{ \pa X_{12}^{ \dot{\s} \s}} \, \cT_{\cA}(X_{12}) \, ,  \label{Three-point building blocks - differential identities 2} \\[2mm]
		\pa_{(2) \, \a \ad} \cT_{\cA}(X_{12}) &= - \frac{1}{x_{23}^{2}} \, \cI_{\a}{}^{\dot{\s}}(x_{23}) \,  \bar{\cI}_{\ad}{}^{\s}(x_{23}) \, \frac{ \pa}{ \pa X_{12}^{ \dot{\s} \s}} \, \cT_{\cA}(X_{12}) \, . \label{Three-point building blocks - differential identities 3}
	\end{align}
\end{subequations}
Now concerning three-point correlation functions, let $\F$, $\J$, $\P$ be primary fields with scale dimensions $\D_{1}$, $\D_{2}$ and $\D_{3}$ respectively. The three-point function may be constructed using the general ansatz
\begin{align}
	\langle \F_{\cA_{1}}(x_{1}) \, \J_{\cA_{2}}(x_{2}) \, \P_{\cA_{3}}(x_{3}) \rangle = \frac{ \cI^{(1)}{}_{\cA_{1}}{}^{\bar{\cA}_{1}}(x_{13}) \,  \cI^{(2)}{}_{\cA_{2}}{}^{\bar{\cA}_{2}}(x_{23}) }{(x_{13}^{2})^{\D_{1}} (x_{23}^{2})^{\D_{2}}}
	\; \cH_{\bar{\cA}_{1} \bar{\cA}_{2} \cA_{3}}(X_{12}) \, , \label{Three-point function - general ansatz}
\end{align} 
where the tensor $\cH_{\bar{\cA}_{1} \bar{\cA}_{2} \cA_{3}}$ encodes all information about the correlation function, and is highly constrained by the conformal symmetry as follows:
\begin{enumerate}
	\item[\textbf{(i)}] Under scale transformations of Minkowski space $x^{m} \mapsto x'^{m} = \l^{-2} x^{m}$, the three-point building blocks transform as $X^{m} \mapsto X'^{m} = \l^{2} X^{m}$. As a consequence, the correlation function transforms as 
	\begin{equation}
		\langle \F_{\cA_{1}}(x_{1}') \, \J_{\cA_{2}}(x_{2}') \, \P_{\cA_{3}}(x_{3}') \rangle = (\l^{2})^{\D_{1} + \D_{2} + \D_{3}} \langle \F_{\cA_{1}}(x_{1}) \, \J_{\cA_{2}}(x_{2}) \,  \P_{\cA_{3}}(x_{3}) \rangle \, ,
	\end{equation}
	which implies that $\cH$ obeys the scaling property
	\begin{equation}
		\cH_{\bar{\cA}_{1} \bar{\cA}_{2} \cA_{3}}(\l^{2} X) = (\l^{2})^{\D_{3} - \D_{2} - \D_{1}} \, \cH_{\bar{\cA}_{1} \bar{\cA}_{2} \cA_{3}}(X) \, , \hspace{5mm} \forall \l \in \mathbb{R} \, \backslash \, \{ 0 \} \, .
	\end{equation}
	This guarantees that the correlation function transforms correctly under scale transformations.
	
	\item[\textbf{(ii)}] If any of the fields $\F$, $\J$, $\P$ obey differential equations, such as conservation laws in the case of conserved current multiplets, then the tensor $\cH$ is also constrained by differential equations. Such constraints may be derived with the aid of identities \eqref{Three-point building blocks - differential identities 2}, \eqref{Three-point building blocks - differential identities 3}.
	
	\item[\textbf{(iii)}] If any (or all) of the operators $\F$, $\J$, $\P$ coincide, the correlation function possesses symmetries under permutations of spacetime points, e.g.
	\begin{equation}
		\langle \F_{\cA_{1}}(x_{1}) \, \F_{\cA_{2}}(x_{2}) \, \P_{\cA_{3}}(x_{3}) \rangle = (-1)^{\e(\F)} \langle \F_{\cA_{2}}(x_{2}) \, \F_{\cA_{1}}(x_{1}) \, \P_{\cA_{3}}(x_{3}) \rangle \, ,
	\end{equation}
	where $\e(\F)$ is the Grassmann parity of $\F$. As a consequence, the tensor $\cH$ obeys constraints which will be referred to as ``point-switch identities". Similar relations may also be derived for two fields which are related by complex conjugation.
	
\end{enumerate}

The constraints above fix the functional form of $\cH$ (and therefore the correlation function) up to finitely many independent parameters. Hence, using the general formula \eqref{H ansatz}, the problem of computing three-point correlation functions is reduced to deriving the general structure of the tensor $\cH$ subject to the above constraints.	

\subsection{Comments regarding differential constraints}\label{subsubsection2.3.1}

An important aspect of this construction which requires further elaboration is that it is sensitive to the configuration of the fields in the correlation function. Indeed, depending on the exact way in which one constructs the general ansatz \eqref{H ansatz}, it can be difficult to impose conservation equations on one of the three fields due to a lack of useful identities such as \eqref{Three-point building blocks - differential identities 2}, \eqref{Three-point building blocks - differential identities 3}. To illustrate this more clearly, consider the following example; suppose we want to determine the solution for the correlation function $\langle \F_{\cA_{1}}(x_{1}) \, \J_{\cA_{2}}(x_{2}) \, \P_{\cA_{3}}(x_{3}) \rangle$, with the ansatz
\begin{equation} \label{H ansatz}
	\langle \F_{\cA_{1}}(x_{1}) \, \J_{\cA_{2}}(x_{2}) \, \P_{\cA_{3}}(x_{3}) \rangle = \frac{ \cI^{(1)}{}_{\cA_{1}}{}^{\bar{\cA}_{1}}(x_{13}) \,  \cI^{(2)}{}_{\cA_{2}}{}^{\bar{\cA}_{2}}(x_{23}) }{(x_{13}^{2})^{\D_{1}} (x_{23}^{2})^{\D_{2}}}
	\; \cH_{\bar{\cA}_{1} \bar{\cA}_{2} \cA_{3}}(X_{12}) \, . 
\end{equation} 
All information about this correlation function is encoded in the tensor $\cH$, however, this particular formulation of the problem prevents us from imposing conservation on the field $\P$ in a straightforward way. To rectify this issue we reformulate the ansatz with $\P$ at the front
\begin{equation} \label{Htilde ansatz}
	\langle \P_{\cA_{3}}(x_{3}) \, \J_{\cA_{2}}(x_{2}) \, \F_{\cA_{1}}(x_{1}) \rangle = \frac{ \cI^{(3)}{}_{\cA_{3}}{}^{\bar{\cA}_{3}}(x_{31}) \,  \cI^{(2)}{}_{\cA_{2}}{}^{\bar{\cA}_{2}}(x_{21}) }{(x_{31}^{2})^{\D_{3}} (x_{21}^{2})^{\D_{2}}}
	\; \tilde{\cH}_{\bar{\cA}_{3} \bar{\cA}_{2} \cA_{1}}(X_{32}) \, . 
\end{equation} 
In this case, all information about this correlation function is now encoded in the tensor $\tilde{\cH}$, which is a completely different solution compared to $\cH$. Conservation on $\P$ can now be imposed by treating $x_{3}$ as the first point with the aid of identities analogous to \eqref{Three-point building blocks - differential identities}, \eqref{Three-point building blocks - differential identities 2}, \eqref{Three-point building blocks - differential identities 3}. What we now need is a simple equation relating the tensors $\cH$ and $\tilde{\cH}$, which correspond to different representations of the same correlation function. If we have equality between the two ansatz above, after some manipulations we obtain the following relation:
\begin{align} \label{Htilde and H relation}
	\tilde{\cH}_{\bar{\cA}_{3} \bar{\cA}_{2} \cA_{1}}(X_{32}) &= (-1)^{\e} \, (x_{13}^{2})^{\D_{3} - \D_{1}} \bigg(\frac{x_{21}^{2}}{x_{23}^{2}} \bigg)^{\hspace{-1mm} \D_{2}} \, \cI^{(1)}{}_{\cA_{1}}{}^{\bar{\cA}_{1}}(x_{13}) \, \bar{\cI}^{(2)}{}_{\bar{\cA}_{2}}{}^{\cA'_{2}}(x_{12}) \,  \cI^{(2)}{}_{\cA'_{2}}{}^{\bar{\cA}'_{2}}(x_{23}) \nonumber \\[-2mm]
	& \hspace{50mm} \times \bar{\cI}^{(3)}{}_{\bar{\cA}_{3}}{}^{\cA_{3}}(x_{13}) \, \cH_{\bar{\cA}_{1} \bar{\cA}'_{2} \cA_{3}}(X_{12}) \, . 
\end{align}
where $\e$ is either $0$ or $1$ depending on the Grassmann parity of the fields $\F$, $\J$, $\P$; since the overall sign is somewhat irrelevant for the purpose of this calculation we will absorb it into the overall sign of $\tilde{\cH}$. In general, this equation is quite impractical to work with due to the presence of both two- and three-point functions, hence, further simplification is required. Let us now introduce some useful definitions; suppose $\cH(X)$ (with indices suppressed) is composed out of a finite basis of linearly independent tensor structures $P_{i}(X)$, i.e $\cH(X) = \sum_{i} a_{i} P_{i}(X)$ where $a_{i}$ are constant complex parameters. We define $\bar{\cH}(X) = \sum_{i} \bar{a}_{i} \bar{P}_{i}(X)$, the conjugate of $\cH$, and also $\cH^{c}(X) = \sum_{i} a_{i} \bar{P}_{i}(X)$, which we will call the complement of $\cH$. As a consequence of \eqref{Inversion tensor identities - spinor case}, the following relation holds:
\begin{align} \label{Hc and H relation}
	\cH^{c}_{\cA_{1} \cA_{2} \bar{\cA}_{3}}(X_{32}) &= (x_{13}^{2} X_{32}^{2})^{\D_{3} - \D_{2} - \D_{1}} \cI^{(1)}{}_{\cA_{1}}{}^{\bar{\cA}_{1}}(x_{13}) \, \cI^{(2)}{}_{\cA_{2}}{}^{\bar{\cA}_{2}}(x_{13}) \nonumber \\
	& \hspace{45mm} \times \bar{\cI}^{(3)}{}_{\bar{\cA}_{3}}{}^{\cA_{3}}(x_{13}) \, \cH_{\bar{\cA}_{1} \bar{\cA}_{2} \cA_{3}}(X_{12}) \, .
\end{align}
This equation is an extension of (2.14) in \cite{Osborn:1993cr} to the spinor representation, and it allows us to construct an equation relating different representations of the same correlation function. After inverting this identity and substituting it directly into \eqref{Htilde and H relation}, we apply identities such as \eqref{Inversion tensor identities - spinor case} to obtain an equation relating $\cH^{c}$ and $\tilde{\cH}$
\begin{equation} \label{Htilde and Hc relation}
	\tilde{\cH}_{\bar{\cA}_{3} \bar{\cA}_{2} \cA_{1}}(X) = (X^{2})^{\D_{1} - \D_{3}} \, \bar{\cI}^{(2)}{}_{\bar{\cA}_{2}}{}^{\cA_{2}}(X) \, \cH^{c}_{\cA_{1} \cA_{2} \bar{\cA}_{3}}(X) \, . 
\end{equation}
It is important to note that this is now an equation in terms of a single variable, $X$, which vastly simplifies the calculations. Indeed, once $\tilde{\cH}$ is obtained we can then impose conservation on $\Pi$ as if it were located at the ``first point''. However, as we will see in the subsequent examples, this transformation is quite difficult to carry out for correlation functions of higher-spin primary operators due to the proliferation of tensor indices.

To summarise, in order to successfully impose all the relevant constraints on the fields in the correlator, we will adhere to the following three step approach:
\begin{enumerate}
	\item Using ansatz \eqref{H ansatz}, construct a solution for $\cH$ that is consistent with the algebraic/tensorial symmetry properties of the fields $\F$, $\J$ and $\P$.
	\item Impose conservation equations on the first and second point using identities \eqref{Three-point building blocks - differential identities}, \eqref{Three-point building blocks - differential identities 2} and \eqref{Three-point building blocks - differential identities 3} to constrain the functional form of the tensor $\cH$.
	\item Reformulate the correlation function using ansatz \eqref{Htilde ansatz}, which allows one to find an explicit relation for $\tilde{\cH}$ in terms of $\cH^{c}$. Conservation of $\P$ may now be imposed as if it were located at the first point.
\end{enumerate}
\subsection{Generating function formalism}\label{subsection2.4}

To study and impose constraints on correlation functions of primary fields with general spins it is often advantageous to use the formalism of generating functions to streamline the calculations. Suppose we must analyse the constraints on a general spin-tensor $\cH_{\cA_{1} \cA_{2} \cA_{3}}(X)$, where $\cA_{1} = \{ \a(i_{1}), \ad(j_{1}) \}, \cA_{2} = \{ \b(i_{2}), \bd(j_{2}) \}, \cA_{3} = \{ \g(i_{3}), \gd(j_{3}) \}$ represent sets of totally symmetric spinor indices associated with the fields at points $x_{1}$, $x_{2}$ and $x_{3}$ respectively. We introduce sets of commuting auxiliary spinors for each point; $U = \{ u, \bar{u} \}$ at $x_{1}$, $ V = \{ v, \bar{v} \}$ at $x_{2}$, and $W = \{ w, \bar{w} \}$ at $x_{3}$, where the spinors satisfy 
\begin{align}
u^2 &= \varepsilon_{\a \b} \, u^{\a} u^{\b}=0\,, & \bar{u}^2& = \varepsilon_{\ad \bd} \, \bar{u}^{\ad} \bar{u}^{\bd}=0\,,  &
v^{2} &= \bar{v}^{2} = 0\,, & w^{2} &= \bar{w}^{2} = 0\,. 
\label{extra1}
\end{align}
Now if we define the objects
\begin{subequations}
	\begin{align}
		\mathbf{U}^{\cA_{1}} &\equiv \mathbf{U}^{\a(i_{1}) \ad(j_{1})} = u^{\a_{1}} \dots u^{\a_{i_{1}}} \bar{u}^{\ad_{1}} \dots \bar{u}^{\ad_{j_{1}}} \, , \\
		\mathbf{V}^{\cA_{2}} &\equiv \mathbf{V}^{\b(i_{2}) \bd(j_{2})} = v^{\b_{1}} \dots v^{\b_{i_{2}}} \bar{v}^{\bd_{1}} \dots \bar{v}^{\bd_{j_{2}}} \, , \\
		\mathbf{W}^{\cA_{3}} &\equiv \mathbf{W}^{\g(i_{3}) \gd(j_{3})} = w^{\g_{1}} \dots w^{\g_{i_{3}}} \bar{w}^{\gd_{1}} \dots \bar{w}^{\gd_{j_{3}}} \, ,
	\end{align}
\end{subequations}
then the generating polynomial for $\cH$ is constructed as follows:
\begin{equation} \label{H - generating polynomial}
	\cH(X; U, V, W) = \cH_{ \cA_{1} \cA_{2} \cA_{3} }(X) \, \mathbf{U}^{\cA_{1}} \mathbf{V}^{\cA_{2}} \mathbf{W}^{\cA_{3}} \, . \\
\end{equation}
%
There is in fact a one-to-one mapping between the space of symmetric traceless spin tensors and the polynomials constructed using the above method. The tensor $\cH$ can then be extracted from the polynomial by acting on it with the following partial derivative operators:
\begin{subequations}
	\begin{align}
		\frac{\pa}{\pa \mathbf{U}^{\cA_{1}} } &\equiv \frac{\pa}{\pa \mathbf{U}^{\a(i_{1}) \ad(j_{1})} } = \frac{1}{i_{1}!j_{1}!} \frac{\pa}{\pa u^{\a_{1}} } \dots \frac{\pa}{\pa u^{\a_{i_{1}}}} \frac{\pa}{\pa \bar{u}^{\ad_{1}}} \dots \frac{\pa }{\pa \bar{u}^{\ad_{j_{1}}}} \, , \\
		\frac{\pa}{\pa \mathbf{V}^{\cA_{2}} } &\equiv \frac{\pa}{\pa \mathbf{V}^{\b(i_{2}) \bd(j_{2})} } = \frac{1}{i_{2}!j_{2}!} \frac{\pa}{\pa v^{\b_{1}} } \dots \frac{\pa}{\pa v^{\b_{i_{2}}}} \frac{\pa}{\pa \bar{v}^{\bd_{1}}} \dots \frac{\pa }{\pa \bar{v}^{\bd_{j_{2}}}} \, , \\
		\frac{\pa}{\pa \mathbf{W}^{\cA_{3}} } &\equiv \frac{\pa}{\pa \mathbf{W}^{\g(i_{3}) \gd(j_{3})} } = \frac{1}{i_{3}!j_{3}!} \frac{\pa}{\pa w^{\g_{1}} } \dots \frac{\pa}{\pa w^{\g_{i_{3}}}} \frac{\pa}{\pa \bar{w}^{\gd_{1}}} \dots \frac{\pa }{\pa \bar{w}^{\gd_{j_{3}}}} \, . 
	\end{align}
\end{subequations}
The tensor $\cH$ is then extracted from the polynomial as follows:
\begin{equation}
	\cH_{\cA_{1} \cA_{2} \cA_{3}}(X) = \frac{\pa}{ \pa \mathbf{U}^{\cA_{1}} } \frac{\pa}{ \pa \mathbf{V}^{\cA_{2}}} \frac{\pa}{ \pa \mathbf{W}^{\cA_{3}} } \, \cH(X; U, V, W) \, .
\end{equation}
Let us point out that methods based on using auxiliary vectors/spinors to create a polynomial are widely used 
in the construction of correlation functions throughout the literature (see e.g.~\cite{Giombi:2011rz, Costa:2011mg, Stanev:2012nq, Zhiboedov:2012bm, Nizami:2013tpa, Elkhidir:2014woa}). However, usually the entire correlator is contracted with auxiliary variables and as a result one produces a polynomial 
depending on all three spacetime points and the auxiliary spinors. In our approach, however, we contract the auxiliary spinors with the tensor $\cH_{ \cA_{1} \cA_{2} \cA_{3} }(X)$, which depends on only a single variable.

Our approach proves to be essential in the construction of correlation functions of higher-spin operators. It also proves to be more computationally tractable, as the polynomial $\cH$, \eqref{H - generating polynomial}, is now constructed out of scalar combinations of $X$, and the auxiliary spinors $U$, $V$ and $W$ with the appropriate homogeneity. Such a polynomial can be constructed out of the following scalar basis structures:
\begin{subequations} \label{Basis scalar structures}
	\begin{align} \label{Basis scalar structures a}
		uv &= u^{\a} v_{\a} \, , & uw &= u^{\a} w_{\a} \, , & vw &= v^{\a} w_{\a} \, , \\
		\bar{u} \bar{v} &= \bar{u}^{\ad} \bar{v}_{\ad} \, , & \bar{u} \bar{w} &= \bar{u}^{\ad} \bar{w}_{\ad} \, , & \bar{v} \bar{w} &= \bar{v}^{\ad} \bar{w}_{\ad} \, , \\
		u X \bar{u} &= u^{\a} \hat{X}_{\a \ad} \bar{u}^{\ad} \, , & u X \bar{v} &= u^{\a} \hat{X}_{\a \ad} \bar{v}^{\ad} \, , & u X \bar{w} &= u^{\a} \hat{X}_{\a \ad} \bar{w}^{\ad} \, , \\
		v X \bar{u} &= v^{\a} \hat{X}_{\a \ad} \bar{u}^{\ad} \, , & v X \bar{v} &= v^{\a} \hat{X}_{\a \ad} \bar{v}^{\ad} \, , & v X \bar{w} &= v^{\a} \hat{X}_{\a \ad} \bar{w}^{\ad} \, , \\
		w X \bar{u} &= w^{\a} \hat{X}_{\a \ad} \bar{u}^{\ad} \, , & w X \bar{v} &= w^{\a} \hat{X}_{\a \ad} \bar{v}^{\ad} \, , & w X \bar{w} &= w^{\a} \hat{X}_{\a \ad} \bar{w}^{\ad} \, ,
	\end{align}
\end{subequations}
subject to cyclic permutations of linear dependence relations such as
\begin{equation}
	( u X \bar{u} ) (\bar{v} \bar{w}) - ( u X \bar{v} ) (\bar{u} \bar{w}) + (u X \bar{w} ) (\bar{u} \bar{v}) = 0 \, .
\end{equation}
There can be more general linear dependence relations for more complicated combinations of the basis structures \eqref{Basis scalar structures}, however, such relations can be obtained computationally.

In general, it is a non-trivial technical problem to come up with an exhaustive list of possible solutions for the polynomial $\cH$ for a given set of spins. Hence, let us introduce a more convenient labelling scheme for the building blocks \eqref{Basis scalar structures}
\begin{subequations}
	\begin{align} 
		P_{1} &= uv \, , & P_{2} &= uw \, , & P_{3} &= vw \, , \\
		Q_{1} &=  u X \bar{v} \, ,  &  Q_{2} &= u X \bar{w} \, ,  &  Q_{3} &= v X \bar{w} \, , \\
		Z_{1} &= u X \bar{u}  \, , & Z_{2} &= v X \bar{v} \, , & Z_{3} &= w X \bar{w} \, .
	\end{align}
\end{subequations}
Now if we also define the objects
\begin{subequations}
	\begin{align} 
		P(k_{1}, k_{2}, k_{3}) &= P_{1}^{k_{1}} P_{2}^{k_{2}} P_{3}^{k_{3}} \, , \\
		Q(X, r_{1}, r_{2}, r_{3}) &= Q_{1}^{r_{1}} Q_{2}^{r_{2}} Q_{3}^{r_{3}} \, , \\
		Z(X, s_{1}, s_{2}, s_{3}) &= Z_{1}^{s_{1}} Z_{2}^{s_{2}} Z_{3}^{s_{3}} \, ,
	\end{align}
\end{subequations}
then the generating function for the polynomial $\cH(X; U, V, W)$ may be defined as follows:
\begin{align} \label{Generating function}
	\cF(X; \G, U,V,W) &= X^{\D_{3} - \D_{2} - \D_{1}} P(k_{1}, k_{2}, k_{3}) \, \bar{P}(\bar{k}_{1}, \bar{k}_{2}, \bar{k}_{3}) \nonumber \\
	& \hspace{15mm}\times  Q(X, r_{1}, r_{2}, r_{3}) \, \bar{Q}(X, \bar{r}_{1}, \bar{r}_{2}, \bar{r}_{3}) \, Z(X, s_{1}, s_{2}, s_{3}) \, ,
\end{align}
where the non-negative integers, $ \G = \{ k_{i}, \bar{k}_{i}, r_{i}, \bar{r}_{i}, s_{i}\}$, $i=1,2,3$, are solutions to the following linear system:
\begin{subequations} \label{Diophantine equations}
	\begin{align}
		k_{1} + k_{2} + s_{1} + r_{1} + r_{2} &= i_{1} \, , &  \bar{k}_{1} + \bar{k}_{2} + s_{1} + \bar{r}_{1} + \bar{r}_{2} &= j_{1} \, , \\
		k_{1} + k_{3} + s_{2} + \bar{r}_{1} + r_{3} &= i_{2} \, , &  \bar{k}_{1} + \bar{k}_{3} + s_{2} + r_{1} + \bar{r}_{3} &= j_{2} \, , \\
		k_{2} + k_{3} + s_{3} + \bar{r}_{2} + \bar{r}_{3} &= i_{3} \, , &  \bar{k}_{2} + \bar{k}_{3} + s_{3} + r_{2} + r_{3} &= j_{3} \, ,
	\end{align}
\end{subequations}
and $i_{1}, i_{2}, i_{3}$, $j_{1}, j_{2}, j_{3}$ are fixed integers which specify the spin-structure of the correlation function. These equations are obtained by comparing the homogeneity of the auxiliary spinors $u$, $\bar{u}$ etc. in the generating function \eqref{Generating function}, against the index structure of the tensor $\cH$. Let us assume there exists a finite number of solutions $\G_{I}$, $I = 1, ..., N$ to \eqref{Diophantine equations} for a given choice of $i_{1}, i_{2}, i_{3}, j_{1}, j_{2}, j_{3}$. Then the most general ansatz for the polynomial $\cH$ in \eqref{H - generating polynomial} is as follows:
\begin{equation}
	\cH(X; U, V, W) = \sum_{I=1}^{N} a_{I} \cF(X; \G_{I}, U, V, W) \, ,
\end{equation}
where $a_{I}$ are a set of complex constants. Hence, constructing the most general ansatz for the generating polynomial $\cH$ is now equivalent to finding all non-negative integer solutions $\G_{I}$ of \eqref{Diophantine equations}, where $i_{1}, i_{2}, i_{3}$ and $j_{1}, j_{2}, j_{3}$ are arbitrary non-negative integers. The solutions correspond to a linearly dependent basis of possible structures in which the polynomial $\cH$ can be decomposed. Using computational methods, we can generate all possible solutions to \eqref{Diophantine equations} for fixed (and in some cases arbitrary) values of the spins. 

In the remaining sections of this paper we will construct solutions for the three-point functions of the fermionic current field $Q_{\a(2k) \ad}$ with the vector current and the energy momentum tensor using the formalism outlined above. We use a combination of the method of systematic decomposition and the generating function approach to reduce the number of possible linearly dependent structures in each case. We present most of our results in terms of the scalar basis structures \eqref{Basis scalar structures}, however, the generating function \eqref{Generating function} underpins most of the calculations.

\section{Correlator \texorpdfstring{$\langle \bar{Q}_{\a \ad(2k) }(x_{1}) \, Q_{\b(2k) \bd}(x_{2}) \, V_{\g \gd}(x_{3}) \rangle$}{< Qb Q V >}}\label{section3}

In this section we will compute the correlation function $\langle \bar{Q} Q V \rangle$, where $V$ is a conserved vector field $V_{\g \gd}$ with scale dimension $3$. The ansatz for this correlator consistent with the general results of subsection \ref{subsection2.3} is
\begin{align} \label{QbQV ansatz}
	\langle \bar{Q}_{\a \ad(2k)}(x_{1}) \, Q_{\b(2k) \bd}(x_{2}) \, V_{\g \gd}(x_{3}) \rangle 
	&= \frac{1}{(x_{13}^{2} x_{23}^{2})^{k + \frac{5}{2}}} \, \cI_{\a}{}^{\ad'}(x_{13}) \, \bar{\cI}_{\ad(2k)}{}^{\a'(2k)}(x_{13}) \nonumber \\[-2mm] & \hspace{25mm} \times \cI_{\b(2k)}{}^{\bd'(2k)}(x_{23}) \, \bar{\cI}_{\bd}{}^{\b'}(x_{23}) \nonumber \\ & \hspace{25mm} \times \cH_{ \a'(2k)  \ad' , \b' \bd'(2k), \g \gd }(X_{12}) \, , 
\end{align}
where $\cH$ is a homogeneous tensor field of degree $q = 3 - 2(k+\tfrac{5}{2}) = - 2 (k+1) $. It is constrained as follows:
\begin{enumerate}
	\item[\textbf{(i)}] Under scale transformations of spacetime $x^{m} \mapsto x'^{m} = \l^{-2} x^{m}$ the three-point building blocks transform as $X^{m} \mapsto X'^{m} = \l^{2} X^{m}$. As a consequence, the correlation function transforms as 
	\begin{equation}
		\langle \bar{Q}_{\a \ad(2k)}(x'_{1}) \, Q_{\b(2k) \bd}(x'_{2}) \, V_{\g \gd}(x'_{3}) \rangle = (\l^{2})^{2k+8} \langle \bar{Q}_{\a \ad(2k)}(x_{1}) \, Q_{\b(2k) \bd}(x_{2}) \, V_{\g \gd}(x_{3}) \rangle \, ,
	\end{equation}
	which implies that $\cH$ obeys the scaling property
	\begin{equation}
		\cH_{ \a(2k) \ad , \b \bd(2k), \g \gd }(\l^{2} X) = (\l^{2})^{q} \, \cH_{ \a(2k) \ad , \b \bd(2k), \g \gd }(X) \, , \hspace{5mm} \forall \l \in \mathbb{R} \, \backslash \, \{ 0 \} \, .
	\end{equation}
	This guarantees that the correlation function transforms correctly under scale transformations.
	
	\item[\textbf{(ii)}] The conservation of the fields $Q$ at $x_{1}$ and $x_{2}$ imply the following constraints on the correlation function:
	\begin{subequations}
		\begin{align}
			\pa_{(1)}^{\ad \a} \langle \bar{Q}_{\a \ad \ad(2k-1)}(x_{1}) \, Q_{\b(2k) \bd}(x_{2}) \, V_{\g \gd}(x_{3}) \rangle = 0 \, , \\
			\pa_{(2)}^{\bd \b} \langle \bar{Q}_{\a \ad(2k)}(x_{1}) \, Q_{\b(2k-1) \b \bd}(x_{2}) \, V_{\g \gd}(x_{3}) \rangle = 0 \, .
		\end{align}
	\end{subequations}
	Using identities \eqref{Three-point building blocks - differential identities 2}, \eqref{Three-point building blocks - differential identities 3} we obtain the following differential constraints on the tensor $\cH$:
	\begin{subequations}
		\begin{align}
			\pa_{X}^{\ad \a} \cH_{ \a(2k-1) \a \ad , \b \bd(2k), \g \gd }(X) = 0 \, , \label{QbQV differential constraint 1a} \\
			\pa_{X}^{\bd \b} \cH_{ \a(2k) \ad, \b \bd \bd(2k-1), \g \gd }(X) = 0 \, , \label{QbQV differential constraint 2a}
		\end{align}
	\end{subequations}
	where $\pa_{X}^{\ad \a} = (\tilde{\s}^{a})^{\ad \a} \frac{\pa}{\pa X^{a}}$. There is also a third constraint equation arising from conservation of $V$ at $x_{3}$,
	\begin{equation}
		\pa_{(3)}^{\gd \g} \langle \bar{Q}_{\a \ad(2k)}(x_{1}) \, Q_{\b(2k) \bd}(x_{2}) \, V_{\g \gd}(x_{3}) \rangle = 0 \, ,
	\end{equation}
	however, there are no identities analogous to \eqref{Three-point building blocks - differential identities 2}, \eqref{Three-point building blocks - differential identities 3} that allow the partial derivative operator acting on $x_{3}$ to pass through the prefactor of \eqref{QQV ansatz}, hence, we use the procedure outlined in subsection \eqref{subsubsection2.3.1}. First we construct an alternative ansatz with $V$ at the front as follows:
	\begin{align}
		\langle  V_{\g \gd}(x_{3}) \, Q_{\b(2k) \bd}(x_{2}) \, \bar{Q}_{\a \ad(2k)}(x_{1}) \rangle &= \frac{1}{(x_{31}^{2})^{3} (x_{21}^{2})^{k + \frac{5}{2}}} \, \cI_{\g}{}^{\gd'}(x_{31}) \, \bar{\cI}_{\gd}{}^{\g'}(x_{31}) \, \nonumber \\[-2mm]
		& \hspace{30mm} \times \cI_{\b(2k)}{}^{\bd'(2k)}(x_{21}) \, \bar{\cI}_{\bd}{}^{\b'}(x_{21}) \nonumber \\ & \hspace{30mm} \times \tilde{\cH}_{ \g' \gd' , \b' \bd'(2k), \a \ad(2k) }(X_{32}) \, .
	\end{align}
	Since the correlation function possesses the following property:
	\begin{equation}
		\langle  V_{\g \gd}(x_{3}) \, Q_{\b(2k) \bd}(x_{2}) \, \bar{Q}_{\a \ad(2k)}(x_{1}) \rangle = - \langle \bar{Q}_{\a \ad(2k)}(x_{1}) \, Q_{\b(2k) \bd}(x_{2}) \, V_{\g \gd}(x_{3}) \rangle \, ,
	\end{equation}
	we can now compute $\tilde{\cH}$ in terms of $\cH$. After some manipulations one finds the following relation:
	\begin{align} \label{QbQV Htilde and H relation}
		\tilde{\cH}_{ \g \gd , \b \bd(2k), \a \ad(2k) }(X_{32}) &= x_{13}^{6} X_{12}^{2k+5} \cI_{\g}{}^{\gd'}(x_{31}) \, \bar{\cI}_{\gd}{}^{\g'}(x_{31}) \, \cI_{\a}{}^{\ad'}(x_{13}) \, \bar{\cI}_{\ad(2k)}{}^{\a'(2k)}(x_{13}) \nonumber \\
		\hspace{10mm} & \times \cI_{\b \dot{\mu}}(x_{13}) \, \bar{\cI}^{ \dot{\mu} \b'}(X_{12}) \, \cI_{\mu(2k) \bd(2k)}(x_{13}) \, \bar{\cI}^{\bd'(2k) \mu(2k) }(X_{12}) \nonumber \\
		\hspace{15mm} & \times \cH_{ \a'(2k) \ad' , \b' \bd'(2k) \g' \gd'}(X_{12}) \, . 
	\end{align}
	This is quite impractical to work with due to the presence of both two-point functions and three-point functions, therefore we will make use of the following relation derived from \eqref{Htilde and H relation}:
	\begin{align}
		\cH_{\a(2k) \ad , \b \bd(2k) , \g \gd}(X_{12}) &= (x_{13}^{2} X_{32}^{2})^{q} \, \cI_{\a(2k)}{}^{\ad'(2k)}(x_{13}) \, \bar{\cI}_{\ad}{}^{\a'}(x_{13}) \nonumber \\ 
		& \times \cI_{\b}{}^{\bd'}(x_{13}) \, \bar{\cI}_{\bd(2k)}{}^{\b'(2k)}(x_{13}) \, \cI_{\g}{}^{\gd'}(x_{13}) \, \bar{\cI}_{\gd}{}^{\g'}(x_{13}) \nonumber \\
		& \times \cH^{c}_{ \a' \ad'(2k) , \b'(2k) \bd' , \g' \gd'}(X_{32}) \, .
	\end{align}
	After substituting this relation directly into \eqref{QbQV Htilde and H relation}, and making use of \eqref{Inversion tensor identities - spinor case}, we obtain the following equation:
	\begin{align} \label{QbQV Htilde and Hc relation}
		\tilde{\cH}_{ \g \gd , \b \bd(2k), \a \ad(2k) }(X) = X^{2k-1} \cI_{\b}{}^{\bd'}(X) \, \bar{\cI}_{\bd(2k)}{}^{\b'(2k)}(X) \, \cH^{c}_{ \a \ad(2k) , \b'(2k) \bd', \g \gd }(X) \, . 
	\end{align}
	The equation relating $\tilde{\cH}$ to $\cH^{c}$ is now expressed in terms of a single variable, the building block vector $X$. Conservation on the third point is now equivalent to imposing the following constraint on the tensor $\tilde{\cH}$:
	\begin{align} \label{QbQV differential constraint 3a}
		\pa_{X}^{\gd \g} \tilde{\cH}_{ \g \gd , \b \bd(2k), \a \ad(2k) }(X) = 0 \, . 
	\end{align}

	\item[\textbf{(iii)}] The correlation function is also constrained by the following reality condition:
	\begin{equation}
		\langle \bar{Q}_{\a \ad(2k)}(x_{1}) \, Q_{\b(2k) \bd}(x_{2}) \, V_{\g \gd}(x_{3}) \rangle = \langle \bar{Q}_{\b \bd(2k)}(x_{2}) \, Q_{\a(2k) \ad}(x_{1}) \, V_{\g \gd}(x_{3}) \rangle^{*} \, ,
	\end{equation}
	which implies the following constraint on the tensor $\cH$:
	\begin{equation} \label{QbQV reality condition a}
		\cH_{ \a(2k) \ad , \b \bd(2k), \g \gd }(X) = - \bar{\cH}_{ \b \bd(2k), \a(2k) \ad, \g \gd }(-X) \, . 
	\end{equation}
\end{enumerate}

Hence, we have to solve for the tensor $\cH$ subject to the above constraints. This is technically quite a challenging problem due to the complicated index structure of the tensor $\cH$. Instead we will streamline the calculations by constructing a generating function as outlined in subsection \ref{subsection2.4}. We introduce the commuting auxiliary spinors $u, \bar{u}, v, \bar{v}, w, \bar{w}$, which satisfy $u^{2} = 0, \, \bar{u}^{2} = 0, \, \text{etc}$, and define the generating function for $\cH$ as follows:
\begin{equation}
	\cH(X; u, \bar{u}, v, \bar{v}, w, \bar{w} ) = \cH_{ \a(2k) \ad , \b \bd(2k), \g \gd }(X) \, \mathbf{U}^{\a(2k) \ad} \mathbf{V}^{\b \bd(2k)} \mathbf{W}^{\g \gd} \, .
\end{equation}
%
The tensor $\cH$ is then obtained from the generating polynomial by acting on it with partial derivatives
\begin{equation}
	\cH_{ \a(2k) \ad , \b \bd(2k), \g \gd }(X)  = \frac{\pa}{ \pa \mathbf{U}^{\a(2k) \ad}} \frac{\pa}{ \pa \mathbf{V}^{\b \bd(2k)}} \frac{\pa}{ \pa  \mathbf{W}^{\g \gd}} \, \cH(X; u, \bar{u}, v, \bar{v}, w, \bar{w} )\, .
\end{equation}
%
Again, the generating function approach simplifies the various algebraic and differential constraints on the tensor $\cH$. In particular, the differential constraints \eqref{QbQV differential constraint 1a} and \eqref{QbQV differential constraint 2a} become
\begin{subequations}
	\begin{align}
		\frac{\pa}{\pa X_{\sigma \dot{\sigma}}} \frac{\pa}{\pa u^{\sigma}} \frac{\pa}{\pa \bar{u}^{\dot{\sigma}}} \, \cH(X; u, \bar{u}, v, \bar{v}, w, \bar{w} ) &= 0 \, , \label{QbQV differential constraint 1b}\\
		\frac{\pa}{\pa X_{\sigma \dot{\sigma}}} \frac{\pa}{\pa v^{\sigma}} \frac{\pa}{\pa \bar{v}^{\dot{\sigma}}} \, \cH(X; u, \bar{u}, v, \bar{v}, w, \bar{w} ) &= 0 \label{QbQV differential constraint 2b} \, ,
	\end{align}
\end{subequations}
while the homogeneity and reality condition \eqref{QbQV reality condition a} become:
\begin{subequations}
	\begin{align}
		\cH(\l^{2} X; u, \bar{u}, v, \bar{v}, w, \bar{w} ) &= (\l^{2})^{q} \cH(X; u, \bar{u}, v, \bar{v}, w, \bar{w} ) \, , \\
		\cH(X; u, \bar{u}, v, \bar{v}, w, \bar{w} ) &= - \bar{\cH}(- X; v, \bar{v}, u, \bar{u}, w, \bar{w} ) \, . \label{QbQV reality condition b}
	\end{align}
\end{subequations}
Our task now is to construct the general solution for the polynomial $\cH$ consistent with the above constraints. 

The general expansion for the polynomial $\cH$ is formed out of products of the basis objects introduced in \eqref{Basis scalar structures}. Let us start by decomposing the polynomial $\cH$, we have:
\begin{equation}
	\cH(X; u, \bar{u}, v, \bar{v}, w, \bar{w} ) = \frac{1}{X^{2k+2}} \, w^{\a} \bar{w}^{\ad} \cF_{\a \ad}(X; u, \bar{u}, v, \bar{v} ) \, ,
\end{equation}
where have used the fact that $\cH$ is homogeneous degree 1 in both $w$ and $\bar{w}$. The vector object $\cF$ is now homogeneous degree 0 in $X$, homogeneous degree 1 in $\bar{u}$, $v$, and homogeneous degree $2k$ in $u$, $\bar{v}$. It may be decomposed further by introducing the following basis vector structures:
\begin{subequations} \label{Basis vector structures}
	\begin{align}
		\cZ_{1 , \, \a \ad} &= \hat{X}_{\a \ad} \, , & \cZ_{2 , \, \a \ad} &= u_{\a} \bar{u}_{\ad} \, , & \cZ_{3 , \, \a \ad} &= u_{\a} \bar{v}_{\ad} \, , \\
		\cZ_{4 , \, \a \ad} &= v_{\a} \bar{u}_{\ad}  \, , &  \cZ_{5 , \, \a \ad} &= v_{\a} \bar{v}_{\ad}  \, .
	\end{align}
\end{subequations}
We then have
\begin{equation}
	\cF_{\a \ad}(X; u, \bar{u}, v, \bar{v}) = \sum_{i=1}^{5} \cZ_{i , \, \a \ad} \, \cF_{i}(X; u, \bar{u}, v, \bar{v} ) \, ,
\end{equation}
where the $\cF_{i}$ are polynomials that are homogeneous degree 0 in $X$, with the appropriate homogeneity in $u, \bar{u}, v, \bar{v}$. It is not too difficult to construct all possible polynomial structures for each $\cZ_{i , \, \a \ad}$:
\begin{subequations}
\begin{itemize}
	\item[] $\cZ_{1}$ structures:
	\begin{align}
		\cF_{1}(X; u, \bar{u}, v, \bar{v}) &= a_{1} (v X \bar{u})(u X \bar{v})^{2k} + a_{2} (u X \bar{u})(v X \bar{v})(u X \bar{v})^{2k-1} \nonumber \\
		& \hspace{15mm} + a_{3} (uv)(\bar{u} \bar{v})(u X \bar{v})^{2k-1} \, ,
	\end{align}
	\item[] $\cZ_{2}$ structures:
	\begin{equation}
		\cF_{2}(X; u, \bar{u}, v, \bar{v}) = a_{4} (v X \bar{v})(u X \bar{v})^{2k-1} \, ,
	\end{equation}
	\item[] $\cZ_{3}$ structures:
	\begin{align}
		\cF_{3}(X; u, \bar{u}, v, \bar{v}) &= a_{5} (v X \bar{u})(u X \bar{v})^{2k-1} + a_{6} (u X \bar{u})(v X \bar{v})(u X \bar{v})^{2k-2} \nonumber \\
		& \hspace{15mm} + a_{7} (uv)(\bar{u} \bar{v})(u X \bar{v})^{2k-2} \, ,
	\end{align}
	\item[] $\cZ_{4}$ structures:
	\begin{equation}
		\cF_{4}(X; u, \bar{u}, v, \bar{v}) = a_{8} (u X \bar{v})^{2k} \, ,
	\end{equation}
	\item[] $\cZ_{5}$ structures:
	\begin{equation}
		\cF_{5}(X; u, \bar{u}, v, \bar{v}) = a_{9} (u X \bar{u})(u X \bar{v})^{2k-1} \, .
	\end{equation}
\end{itemize}
\end{subequations}
However, not all of these structures are linearly independent. In particular it may be shown that $\cF_{\a \ad}(X; u, \bar{u}, v, \bar{v}) = 0$ for the choice
\begin{align}
	a_{2} &= - a_{1} \, ,  &  a_{3} &= a_{1} \, ,  &  a_{i} &= 0 \, , & i &= 4, ... \, ,9 \, .
\end{align}
Therefore we can construct a linearly independent basis of polynomial structures by removing the $a_{1}$ structure, which leaves us with 8 independent structures to consider. We now impose the differential constraints and point switch identities using Mathematica. After imposing \eqref{QbQV differential constraint 1b}, we obtain the following $k$-dependent relations:
\begin{align}
	a_{4} &= \frac{(1-2k) a_{2} + (1+2k) a_{3} }{1+2k} \, , & a_{7} &= \frac{(-1+2k) ( a_{2} + (1+2k) a_{3}) }{2k(1+2k)} \, ,
\end{align}
in addition to $a_{5} = a_{6} = 0$. Next we impose \eqref{QbQV differential constraint 2b}, from which we obtain
\begin{equation}
	a_{9} = \frac{(1-2k) a_{2} + (1+2k) a_{3} }{1+2k} \, .
\end{equation}
Hence, the correlation function is determined up to three independent complex parameters, $a_{2}$, $a_{3}$ and $a_{8}$. We now must impose the reality condition \eqref{QbQV reality condition b}. Using Mathematica, we find that $a_{2} = \text{i} \tilde{a}_{2}$, $a_{3} = \text{i} \tilde{a}_{3}$,  $a_{8} = \text{i} \tilde{a}_{8}$, where $\tilde{a}_{2}$, $\tilde{a}_{3}$ , $\tilde{a}_{8}$ are three real constant parameters. 

It remains to demonstrate that this correlation function is conserved at $x_{3}$ in accordance with conservation of the vector current. First we compute the tensor $\tilde{\cH}$ using \eqref{QbQV Htilde and Hc relation}. This may be written more compactly in the generating function formalism; to do this we introduce the following differential operators:
\begin{align}
	(v X \pa \bar{s}) = v^{\a} \hat{X}_{\a}{}^{\ad} \frac{\pa}{\pa \bar{s}^{\ad}} \, , \hspace{10mm} (\pa s X \bar{v}) =  \frac{\pa}{\pa s^{\a}} \hat{X}^{\a}{}_{\ad} \bar{v}^{\ad}  \, .
\end{align}
The relation \eqref{QbQV Htilde and Hc relation} is now equivalent to
\begin{align}
	\tilde{\cH}(X; w, \bar{w}, v, \bar{v}, u, \bar{u}) = \frac{1}{(2k)!} \, X^{2k-1} (v X \pa \bar{s}) \, (\pa s X \bar{v})^{2k} \, \cH^{c}(X; u, \bar{u}, s, \bar{s}, w, \bar{w}) \, .
\end{align}
Conservation on the third point \eqref{QbQV differential constraint 3a} is equivalent to imposing the following constraint on $\tilde{\cH}$:
\begin{align}
	\frac{\pa}{\pa X_{\sigma \dot{\sigma}}} \frac{\pa}{\pa w^{\sigma}} \frac{\pa}{\pa \bar{w}^{\dot{\sigma}}} \, \tilde{\cH}(X; w, \bar{w}, v, \bar{v}, u, \bar{u} ) = 0 \, .
\end{align}
It may be shown using Mathematica that this is satisfied up to $k=4$. Beyond $k=4$ the calculations for \eqref{QbQV Htilde and Hc relation} seem to become very computationally intensive; however, we have no reason to expect that the result will change for higher values of $k$. Hence, we are reasonably confident that $\langle \bar{Q} Q V \rangle$ is fixed up to three independent real parameters.


\section{Correlator \texorpdfstring{$\langle Q_{\a(2k) \ad }(x_{1}) \, Q_{\b(2k) \bd}(x_{2}) \, V_{\g \gd}(x_{3}) \rangle$}{< Q Q V >}}\label{section4}

In this section we will compute the correlation function $\langle Q Q V \rangle$. The ansatz for this correlator consistent with the general results of subsection \ref{subsection2.3} is
\begin{align} \label{QQV ansatz}
	\langle Q_{\a(2k) \ad}(x_{1}) \, Q_{\b(2k) \bd}(x_{2}) \, V_{\g \gd}(x_{3}) \rangle 
	&= \frac{1}{(x_{13}^{2} x_{23}^{2})^{k + \frac{5}{2}}} \, \cI_{\a(2k)}{}^{\ad'(2k)}(x_{13}) \, \bar{\cI}_{\ad}{}^{\a'}(x_{13}) \nonumber \\[-2mm] & \hspace{25mm} \times \cI_{\b(2k)}{}^{\bd'(2k)}(x_{23}) \, \bar{\cI}_{\bd}{}^{\b'}(x_{23}) \nonumber \\ & \hspace{25mm} \times \cH_{ \a' \ad'(2k) , \b' \bd'(2k), \g \gd }(X_{12}) \, , 
\end{align}
where $\cH$ is a homogeneous tensor field of degree $q = 3 - 2(k+\tfrac{5}{2}) = - 2 (k+1) $. It is constrained as follows:
\begin{enumerate}
	\item[\textbf{(i)}] Under scale transformations of spacetime $x^{m} \mapsto x'^{m} = \l^{-2} x^{m}$ the three-point building blocks transform as $X^{m} \mapsto X'^{m} = \l^{2} X^{m}$. As a consequence, the correlation function transforms as 
	\begin{equation}
		\langle Q_{\a(2k) \ad}(x'_{1}) \, Q_{\b(2k) \bd}(x'_{2}) \, V_{\g \gd}(x'_{3}) \rangle = (\l^{2})^{2k+8} \langle Q_{\a(2k) \ad}(x_{1}) \, Q_{\b(2k) \bd}(x_{2}) \, V_{\g \gd}(x_{3}) \rangle \, ,
	\end{equation}
	which implies that $\cH$ obeys the scaling property
	\begin{equation}
		\cH_{ \a \ad(2k) , \b \bd(2k), \g \gd }(\l^{2} X) = (\l^{2})^{q} \, \cH_{ \a \ad(2k) , \b \bd(2k), \g \gd }(X) \, , \hspace{5mm} \forall \l \in \mathbb{R} \, \backslash \, \{ 0 \} \, .
	\end{equation}
	This guarantees that the correlation function transforms correctly under scale transformations.
	
	\item[\textbf{(ii)}] The conservation of the fields $Q$ at $x_{1}$ and $x_{2}$ imply the following constraints on the correlation function:
	\begin{subequations}
		\begin{align}
			\pa_{(1)}^{\ad \a} \langle Q_{\a(2k-1) \a \ad}(x_{1}) \, Q_{\b(2k) \bd}(x_{2}) \, V_{\g \gd}(x_{3}) \rangle = 0 \, , \\
			\pa_{(2)}^{\bd \b} \langle Q_{\a(2k) \ad}(x_{1}) \, Q_{\b(2k-1) \b \bd}(x_{2}) \, V_{\g \gd}(x_{3}) \rangle = 0 \, .
		\end{align}
	\end{subequations}
	Using identities \eqref{Three-point building blocks - differential identities 2}, \eqref{Three-point building blocks - differential identities 3}, we obtain the following differential constraints on the tensor $\cH$:
	\begin{subequations}
		\begin{align}
			\pa_{X}^{\ad \a} \cH_{ \a \ad \ad(2k-1) , \b \bd(2k), \g \gd }(X) = 0 \, , \label{QQV differential constraint 1a} \\
			\pa_{X}^{\bd \b} \cH_{ \a \ad(2k) , \b \bd \bd(2k-1), \g \gd }(X) = 0 \, . \label{QQV differential constraint 2a}
		\end{align}
	\end{subequations}
	There is also a third constraint equation arising from conservation of $V$ at $x_{3}$:
	\begin{equation}
		\pa_{(3)}^{\gd \g} \langle Q_{\a(2k) \ad}(x_{1}) \, Q_{\b(2k) \bd}(x_{2}) \, V_{\g \gd}(x_{3}) \rangle = 0 \, ,
	\end{equation}
	Similar to the previous example, we use the procedure outlined in subsection \ref{subsubsection2.3.1} and find the following relation between $\cH^{c}$ and $\tilde{\cH}$:
	\begin{align} \label{QQV Htilde and Hc relation}
	\tilde{\cH}_{ \g \gd , \b \bd(2k), \a(2k) \ad }(X) = X^{2k-1} \cI_{\b}{}^{\bd'}(X) \, \bar{\cI}_{\bd(2k)}{}^{\b'(2k)}(X) \, \cH^{c}_{ \a(2k) \ad , \b'(2k) \bd', \g \gd }(X) \, . 
	\end{align}
	Conservation on the third point is now tantamount to imposing the constraint
	\begin{align} \label{QQV differential constraint 3a}
	\pa_{X}^{\gd \g} \tilde{\cH}_{ \g \gd , \b \bd(2k), \a(2k) \ad }(X) = 0 \, . 
	\end{align}

	\item[\textbf{(iii)}] The correlation function possesses the following symmetry property under exchange of the fields at $x_{1}$ and $x_{2}$:
	\begin{equation}
		\langle Q_{\a(2k) \ad}(x_{1}) \, Q_{\b(2k) \bd}(x_{2}) \, V_{\g \gd}(x_{3}) \rangle = - \langle Q_{\b(2k) \bd}(x_{2}) \, Q_{\a(2k) \ad}(x_{1}) \, V_{\g \gd}(x_{3}) \rangle \, .
	\end{equation}
	This implies the following constraint on the tensor $\cH$:
	\begin{equation} \label{QQV point switch a}
		\cH_{ \a \ad(2k) , \b \bd(2k), \g \gd }(X) = - \cH_{ \b \bd(2k), \a \ad(2k), \g \gd }(-X) \, . 
	\end{equation}
\end{enumerate}

Hence, we have to solve for the tensor $\cH$ subject to the above constraints. Analogous to the previous example in section \ref{section3}, we streamline the calculations by constructing a generating function, which is defined as follows:
\begin{equation}
	\cH(X; u, \bar{u}, v, \bar{v}, w, \bar{w} ) = \cH_{ \a \ad(2k) , \b \bd(2k), \g \gd }(X) \, \mathbf{U}^{\a \ad(2k)} \mathbf{V}^{\b \bd(2k)} \mathbf{W}^{\g \gd} \, .
\end{equation}
%
The tensor $\cH$ is then extracted from the generating polynomial by acting on it with partial derivatives
\begin{equation}
 \cH_{ \a \ad(2k) , \b \bd(2k), \g \gd }(X)  = \frac{\pa}{ \pa \mathbf{U}^{\a \ad(2k)}} \frac{\pa}{ \pa \mathbf{V}^{\b \bd(2k)}} \frac{\pa}{ \pa  \mathbf{W}^{\g \gd}} \, \cH(X; u, \bar{u}, v, \bar{v}, w, \bar{w} )\, .
\end{equation}
%
As will be seen shortly, the generating function approach simplifies the various algebraic and differential constraints on the tensor $\cH$. In particular, the differential constraints \eqref{QQV differential constraint 1a} and \eqref{QQV differential constraint 2a} become
\begin{subequations}
	\begin{align}
		\frac{\pa}{\pa X_{\sigma \dot{\sigma}}} \frac{\pa}{\pa u^{\sigma}} \frac{\pa}{\pa \bar{u}^{\dot{\sigma}}} \, \cH(X; u, \bar{u}, v, \bar{v}, w, \bar{w} ) &= 0 \, , \label{QQV differential constraint 1b} \\
		\frac{\pa}{\pa X_{\sigma \dot{\sigma}}} \frac{\pa}{\pa v^{\sigma}} \frac{\pa}{\pa \bar{v}^{\dot{\sigma}}} \, \cH(X; u, \bar{u}, v, \bar{v}, w, \bar{w} ) &= 0 \, , \label{QQV differential constraint 2b}
	\end{align}
\end{subequations}
while the homogeneity and point switch constraints become:
\begin{subequations}
	\begin{align}
		\cH(\l^{2} X; u, \bar{u}, v, \bar{v}, w, \bar{w} ) &= (\l^{2})^{q} \, \cH(X; u, \bar{u}, v, \bar{v}, w, \bar{w} ) \, , \\
		\cH(X; u, \bar{u}, v, \bar{v}, w, \bar{w} ) &= - \cH(- X; v, \bar{v}, u, \bar{u}, w, \bar{w} ) \, . \label{QQV point switch b}
	\end{align}
\end{subequations}
Our task is now to construct the general solution for the polynomial $\cH$ consistent with the above constraints.

The general expansion for the polynomial $\cH$ is then formed out of products of the basis objects above. Let us start by decomposing the polynomial $\cH$, we have:
\begin{equation}
	\cH(X; u, \bar{u}, v, \bar{v}, w, \bar{w} ) = \frac{1}{X^{2k+2}} w^{\a} \bar{w}^{\ad} \cF_{\a \ad}(X; u, \bar{u}, v, \bar{v} ) \, ,
\end{equation}
where have used the fact that $\cH$ is homogeneous degree 1 in both $w$ and $\bar{w}$. The vector object $\cF$ is now homogeneous degree 0 in $X$, degree 1 in $u$, $v$, and degree $2k$ in $\bar{u}$, $\bar{v}$. It may be decomposed further using the structures defined in \eqref{Basis vector structures}:
\begin{equation}
	\cF_{\a \ad}(X; u, \bar{u}, v, \bar{v}) = \sum_{i=1}^{5} \cZ_{i , \, \a \ad} \, \cF_{i}(X; u, \bar{u}, v, \bar{v} ) \, ,
\end{equation}
where the $\cF_{i}$ are polynomials that are homogeneous degree 0 in $X$, with the appropriate homogeneity in $u, \bar{u}, v, \bar{v}$. It is not too difficult to construct all possible polynomial structures for each $\cZ_{i , \, \a \ad}$, we find:
\begin{subequations}
\begin{itemize}
	\item[] $\cZ_{1}$ structures:
	\begin{align}
		\cF_{1}(X; u, \bar{u}, v, \bar{v}) &= a_{1} (uv)(\bar{u}\bar{v})^{2k} + a_{2} (uX\bar{u})(vX\bar{v})(\bar{u}\bar{v})^{2k-1} \nonumber \\
		& \hspace{15mm} + a_{3} (uX\bar{v})(vX\bar{u})(\bar{u}\bar{v})^{2k-1} \, ,
	\end{align}
	\item[] $\cZ_{2}$ structures:
	\begin{equation}
		\cF_{2}(X; u, \bar{u}, v, \bar{v}) = a_{4} (vX\bar{v})(\bar{u}\bar{v})^{2k-1} \, ,
	\end{equation}
	\item[] $\cZ_{3}$ structures:
	\begin{equation}
		\cF_{3}(X; u, \bar{u}, v, \bar{v}) = a_{5} (vX\bar{u})(\bar{u}\bar{v})^{2k-1} \, ,
	\end{equation}
	\item[] $\cZ_{4}$ structures:
	\begin{equation}
		\cF_{4}(X; u, \bar{u}, v, \bar{v}) = a_{6} (uX\bar{v})(\bar{u}\bar{v})^{2k-1} \, ,
	\end{equation}
	\item[] $\cZ_{5}$ structures:
	\begin{equation}
		\cF_{5}(X; u, \bar{u}, v, \bar{v}) = a_{7} (uX\bar{u})(\bar{u}\bar{v})^{2k-1} \, .
	\end{equation}
\end{itemize}
\end{subequations}
However, not all of these structures are linearly independent. In particular it may be shown that $\cF_{\a \ad}(X; u, \bar{u}, v, \bar{v}) = 0$ for the choice
\begin{align}
	a_{3} &= - a_{2} \, ,  &  a_{4} &= - a_{1} + a_{2} \, ,  &  a_{5} &= a_{1} - a_{2} \, , &  a_{6} &= a_{1} - a_{2} \, , &  a_{7} &= - a_{1} + a_{2} \, ,
\end{align}
Therefore we can construct a linearly independent basis of polynomial structures by removing the $a_{1}$ and $a_{2}$ structures, hence, there are only 5 independent structures remaining. We now impose the differential constraints and point switch identities using Mathematica. After imposing \eqref{QQV differential constraint 1b}, we obtain the following $k$-dependent relations between the coefficients:
\begin{align}
	a_{4} &= - \frac{1}{2k} a_{3} \, , &  a_{5} &= \frac{3(1+2k)}{2k(3+2k)} a_{3} \, , & a_{7} &= - \frac{(1+2k)a_{3} + (3+2k) a_{6}}{3 + 8k + 4k^{2}} \, .
\end{align}
At this stage only two independent coefficients remain. Next we impose \eqref{QQV differential constraint 2b}, from which we obtain
\begin{align} \label{QQV coefficient constraints}
	a_{4} &= - \frac{1}{2k} a_{3} \, , &  a_{5} &= \frac{3(1+2k)}{2k(3+2k)} a_{3} \, , & a_{6} &= \frac{3(1+2k)}{2k(3+2k)} a_{3} \, , & a_{7} &= - \frac{1}{2k} a_{3} \, . 
\end{align}
Hence, the correlation function is determined up to a single complex parameter, $a_{3} = a$. However, it may be shown that this solution is not compatible with the point switch identity \eqref{QQV point switch b}, hence, this correlation vanishes in general.

\section{Correlator \texorpdfstring{$\langle \bar{Q}_{\a \ad(2k) }(x_{1}) \, Q_{\b(2k) \bd}(x_{2}) \, T_{\g(2) \gd(2)}(x_{3}) \rangle$}{< Qb Q T >}}\label{section5}

In this section we will compute the correlation function $\langle \bar{Q} Q T \rangle$, where $T$ is the energy momentum tensor $T_{\g(2) \gd(2) }$ with scale dimension $4$. The ansatz for this correlator consistent with the general results of subsection \ref{subsection2.3} is
\begin{align} \label{QbQT ansatz}
	\langle \bar{Q}_{\a \ad(2k)}(x_{1}) \, Q_{\b(2k) \bd}(x_{2}) \, T_{\g(2) \gd(2)}(x_{3}) \rangle 
	&= \frac{1}{(x_{13}^{2} x_{23}^{2})^{k + \frac{5}{2}}} \, \cI_{\a}{}^{\ad'}(x_{13}) \, \bar{\cI}_{\ad(2k)}{}^{\a'(2k)}(x_{13}) \nonumber \\[-2mm] & \hspace{25mm} \times \cI_{\b(2k)}{}^{\bd'(2k)}(x_{23}) \, \bar{\cI}_{\bd}{}^{\b'}(x_{23}) \nonumber \\ & \hspace{25mm} \times \cH_{ \a'(2k)  \ad' , \b' \bd'(2k), \g(2) \gd(2) }(X_{12}) \, , 
\end{align}
where $\cH$ is a homogeneous tensor field of degree $q = 4 - 2(k+\tfrac{5}{2}) = - 2 k - 1 $. It is constrained as follows:
\begin{enumerate}
	\item[\textbf{(i)}] Under scale transformations of spacetime $x^{m} \mapsto x'^{m} = \l^{-2} x^{m}$ the three-point building blocks transform as $X^{m} \mapsto X'^{m} = \l^{2} X^{m}$. As a consequence, the correlation function transforms as 
	\begin{align}
		\langle \bar{Q}_{\a \ad(2k)}(x'_{1}) \, Q_{\b(2k) \bd}(x'_{2}) \, T_{\g(2) \gd(2)}(x'_{3}) \rangle &= \nonumber \\
		& \hspace{-30mm }(\l^{2})^{2k+9} \langle \bar{Q}_{\a \ad(2k)}(x_{1}) \, Q_{\b(2k) \bd}(x_{2}) \, T_{\g(2) \gd(2)}(x_{3}) \rangle \, ,
	\end{align}
	which implies that $\cH$ obeys the scaling property
	\begin{equation}
		\cH_{ \a(2k) \ad , \b \bd(2k), \g(2) \gd(2) }(\l^{2} X) = (\l^{2})^{q} \cH_{ \a(2k) \ad , \b \bd(2k), \g(2) \gd(2) }(X) \, , \hspace{5mm} \forall \l \in \mathbb{R} \, \backslash \, \{ 0 \} \, .
	\end{equation}
	This guarantees that the correlation function transforms correctly under scale transformations.
	
	\item[\textbf{(ii)}] The conservation of the fields $Q$ at $x_{1}$ and $x_{2}$ imply the following constraints on the correlation function:
	\begin{subequations}
		\begin{align}
			\pa_{(1)}^{\ad \a} \langle \bar{Q}_{\a \ad \ad(2k-1)}(x_{1}) \, Q_{\b(2k) \bd}(x_{2}) \, T_{\g(2) \gd(2)}(x_{3}) \rangle = 0 \, , \\
			\pa_{(2)}^{\bd \b} \langle \bar{Q}_{\a \ad(2k)}(x_{1}) \, Q_{\b(2k-1) \b \bd}(x_{2}) \, T_{\g(2) \gd(2)}(x_{3}) \rangle = 0 \, .
		\end{align}
	\end{subequations}
	Using identities \eqref{Three-point building blocks - differential identities 2}, \eqref{Three-point building blocks - differential identities 3} we obtain the following differential constraints on the tensor $\cH$:
	\begin{subequations}
		\begin{align}
			\pa_{X}^{\ad \a} \cH_{ \a(2k-1) \a \ad , \b \bd(2k), \g(2) \gd(2) }(X) = 0 \, , \label{QbQT differential constraint 1a} \\
			\pa_{X}^{\bd \b} \cH_{ \a(2k) \ad, \b \bd \bd(2k-1), \g(2) \gd(2) }(X) = 0 \, , \label{QbQT differential constraint 2a}
		\end{align}
	\end{subequations}
	There is also a third constraint equation arising from conservation of $V$ at $x_{3}$,
	\begin{equation}
		\pa_{(3)}^{\gd \g} \langle \bar{Q}_{\a \ad(2k)}(x_{1}) \, Q_{\b(2k) \bd}(x_{2}) \, T_{\g(2) \gd(2)}(x_{3}) \rangle = 0 \, .
	\end{equation}
	Using the same procedure as the previous examples, we construct an alternative ansatz for the correlation function as follows:
	\begin{align}
		\langle T_{\g(2) \gd(2)}(x_{3}) \, Q_{\b(2k) \bd}(x_{2}) \, \bar{Q}_{\a \ad(2k)}(x_{1}) \rangle &= \frac{1}{(x_{31}^{2})^{3} (x_{21}^{2})^{\frac{7}{2}}} \, \cI_{\g(2)}{}^{\gd'(2)}( x_{31}) \, \bar{\cI}_{\gd(2)}{}^{\g'(2)}( x_{31}) \nonumber \\[-2mm]
		& \hspace{25mm} \times \cI_{\b(2k)}{}^{\bd'(2k)}( x_{21}) \,  \bar{\cI}_{\bd}{}^{\bd'}( x_{21}) \nonumber \\ & \hspace{25mm} \times \tilde{\cH}_{ \g'(2) \gd'(2) , \b' \bd'(2k), \a \ad(2k) }(X_{32}) \, .
	\end{align}
	Now due to the property
	\begin{equation}
		\langle  T_{\g(2) \gd(2)}(x_{3}) \, Q_{\b(2k) \bd}(x_{2}) \, \bar{Q}_{\a \ad(2k)}(x_{1}) \rangle = - \langle \bar{Q}_{\a \ad(2k)}(x_{1}) \, Q_{\b(2k) \bd}(x_{2}) \, T_{\g(2) \gd(2)}(x_{3}) \rangle \, ,
	\end{equation}
	we have a way to compute $\tilde{\cH}$ in terms of $\cH$. After some manipulations we find
	\begin{align} \label{QbQT Htilde and H relation}
		\tilde{\cH}_{ \g(2) \gd(2) , \b \bd(2k), \a \ad(2k) }(X_{32}) &= x_{13}^{8} X_{12}^{2k+5} \cI_{\g(2)}{}^{\gd'(2)}(x_{31}) \, \bar{\cI}_{\gd(2)}{}^{\g'(2)}(x_{31}) \nonumber \\
		& \times \cI_{\a}{}^{\ad'}(x_{13}) \, \bar{\cI}_{\ad(2k)}{}^{\a'(2k)}(x_{13}) \, \cI_{\b \dot{\mu}}(x_{13}) \, \bar{\cI}^{ \dot{\mu} \b'}(X_{12}) \nonumber \\
		& \times \cI_{\mu(2k) \bd(2k)}(x_{13}) \, \bar{\cI}^{\bd'(2k) \mu(2k) }(X_{12}) \nonumber \\
		& \times \cH_{ \a'(2k) \ad' , \b' \bd'(2k), \g'(2) \gd'(2)}(X_{12}) \, . 
	\end{align}
	We now make use of the following identity derived from \eqref{Hc and H relation}:
	\begin{align}
		\cH_{\a(2k) \ad , \b \bd(2k) , \g(2) \gd(2)}(X_{12}) &= (x_{13}^{2} X_{32}^{2})^{q} \, \cI_{\a(2k)}{}^{\ad'(2k)}(x_{13}) \, \bar{\cI}_{\ad}{}^{\a'}(x_{13}) \nonumber \\ & \times \cI_{\b}{}^{\bd'}(x_{13}) \, \bar{\cI}_{\bd(2k)}{}^{\bd'(2k)}(x_{13}) \nonumber \\
		& \times \cI_{\g(2)}{}^{\gd'(2)}(x_{13}) \,  \bar{\cI}_{\gd(2)}{}^{\gd'(2)}(x_{13}) \nonumber \\
		& \times \cH^{c}_{ \a' \ad'(2k) , \b'(2k) \bd' , \g'(2) \gd'(2)}(X_{32}) \, .
	\end{align}
	After substituting this equation into \eqref{QbQT Htilde and H relation}, we obtain the relation
	\begin{align} \label{QbQT Htilde and Hc relation}
		\tilde{\cH}_{ \g(2) \gd(2) , \b \bd(2k), \a \ad(2k) }(X) = X^{2k-3} \cI_{\b}{}^{\bd'}(X) \, \bar{\cI}_{\bd(2k)}{}^{\b'(2k)}(X) \, \cH^{c}_{ \a \ad(2k) , \b'(2k) \bd', \g(2) \gd(2) }(X) \, . 
	\end{align}
	Conservation at $x_{3}$ is now equivalent to imposing the following constraint on the tensor $\tilde{\cH}$:
	\begin{align} \label{QbQT differential constraint 3a}
		\pa_{X}^{\dot{\s} \s} \tilde{\cH}_{ \s \dot{\s} \g \gd , \b \bd(2k), \a \ad(2k) }(X) = 0 \, . 
	\end{align}

	\item[\textbf{(iii)}] The correlation function is also constrained by the reality condition
	\begin{equation}
		\langle \bar{Q}_{\a \ad(2k)}(x_{1}) \, Q_{\b(2k) \bd}(x_{2}) \, T_{\g(2) \gd(2)}(x_{3}) \rangle = \langle \bar{Q}_{\b \bd(2k)}(x_{2}) \, Q_{\a(2k) \ad}(x_{1}) \, T_{\g(2) \gd(2)}(x_{3}) \rangle^{*} \, .
	\end{equation}
	This implies the following constraint on the tensor $\cH$:
	\begin{equation} \label{QbQT reality condition a}
		\cH_{ \a(2k) \ad , \b \bd(2k), \g(2) \gd(2) }(X) = - \bar{\cH}_{ \b \bd(2k), \a(2k) \ad, \g(2) \gd(2) }(-X) \, . 
	\end{equation}
\end{enumerate}

Hence, we have to solve for the tensor $\cH$ subject to the above constraints. Analogous to the previous examples we streamline the calculations by constructing a generating function, which is defined as follows:
\begin{equation}
	\cH(X; u, \bar{u}, v, \bar{v}, w, \bar{w} ) = \cH_{ \a(2k) \ad , \b \bd(2k), \g(2) \gd(2) }(X) \, \mathbf{U}^{\a(2k) \ad} \mathbf{V}^{\b \bd(2k)} \mathbf{W}^{\g(2) \gd(2)} \, .
\end{equation}
%
The tensor $\cH$ is then obtained from the generating polynomial by acting on it with partial derivatives
\begin{equation}
	\cH_{ \a(2k) \ad , \b \bd(2k), \g(2) \gd(2) }(X)  = \frac{\pa}{ \pa \mathbf{U}^{\a(2k) \ad}} \frac{\pa}{ \pa \mathbf{V}^{\b \bd(2k)}} \frac{\pa}{ \pa  \mathbf{W}^{\g(2) \gd(2)}} \, \cH(X; u, \bar{u}, v, \bar{v}, w, \bar{w} )\, .
\end{equation}
%
Again, the generating function approach simplifies the various algebraic and differential constraints on the tensor $\cH$. In particular, the differential constraints \eqref{QbQT differential constraint 1a} and \eqref{QbQT differential constraint 2a} become
\begin{subequations}
	\begin{align}
		\frac{\pa}{\pa X_{\sigma \dot{\sigma}}} \frac{\pa}{\pa u^{\sigma}} \frac{\pa}{\pa \bar{u}^{\dot{\sigma}}} \, \cH(X; u, \bar{u}, v, \bar{v}, w, \bar{w} ) &= 0 \, , \label{QbQT differential constraint 1b} \\
		\frac{\pa}{\pa X_{\sigma \dot{\sigma}}} \frac{\pa}{\pa v^{\sigma}} \frac{\pa}{\pa \bar{v}^{\dot{\sigma}}} \, \cH(X; u, \bar{u}, v, \bar{v}, w, \bar{w} ) &= 0 \, , \label{QbQT differential constraint 2b}
	\end{align}
\end{subequations}
while the homogeneity and point switch constraints become:
\begin{subequations}
	\begin{align}
		\cH(\l^{2} X; u, \bar{u}, v, \bar{v}, w, \bar{w} ) &= (\l^{2})^{q} \, \cH(X; u, \bar{u}, v, \bar{v}, w, \bar{w} ) \, , \\
		\cH(X; u, \bar{u}, v, \bar{v}, w, \bar{w} ) &= - \bar{\cH}(- X; v, \bar{v}, u, \bar{u}, w, \bar{w} ) \, . \label{QbQT reality condition b}
	\end{align}
\end{subequations}
Let us now construct the general solution for the polynomial $\cH$ consistent with the above constraints. We start by decomposing the polynomial $\cH$ as follows:
\begin{equation}
	\cH(X; u, \bar{u}, v, \bar{v}, w, \bar{w} ) = \frac{1}{X^{2k+1}} \, \bar{u}^{\ad} v^{\a} \cF_{\a \ad}(X; u, \bar{v}, w, \bar{w} ) \, ,
\end{equation}
where have used the fact that $\cH$ is homogeneous degree 1 in both $\bar{u}$ and $v$. The vector object $\cF$ is now homogeneous degree 0 in $X$, homogeneous degree 2 in $w$, $\bar{w}$, and homogeneous degree $2k$ in $u$, $\bar{v}$. It may be decomposed further by defining the following basis vectors:
\begin{subequations}
	\begin{align}
		\cZ_{1 , \, \a \ad} &= \hat{X}_{\a \ad} \, , & \cZ_{2 , \, \a \ad} &= u_{\a} \bar{v}_{\ad} \, , & \cZ_{3 , \, \a \ad} &= u_{\a} \bar{w}_{\ad} \, , \\
		\cZ_{4 , \, \a \ad} &= w_{\a} \bar{v}_{\ad}  \, , &  \cZ_{5 , \, \a \ad} &= w_{\a} \bar{w}_{\ad}  \, .
	\end{align}
\end{subequations}
We then have
\begin{equation}
	\cF_{\a \ad}(X; u, \bar{v}, w, \bar{w}) = \sum_{i=1}^{5} \cZ_{i , \, \a \ad} \, \cF_{i}(X; u, \bar{v}, w, \bar{w} ) \, ,
\end{equation}
where the $\cF_{i}$ are polynomials that are homogeneous degree 0 in $X$, with the appropriate homogeneity in $u, \bar{v}, w, \bar{w} $. The complete list of possible polynomial structures for each $\cZ_{i , \, \a \ad}$ is:
\begin{subequations}
\begin{itemize}
	\item[] $\cZ_{1}$ structures:
	\begin{align}
		\cF_{1}(X; u, \bar{v}, w, \bar{w}) &= a_{1} \, (uw)^{2} (\bar{v} \bar{w}) (u X \bar{v})^{2k-2} + a_{2} \, (w X \bar{w})^{2} (u X \bar{v})^{2k} \nonumber \\
		& \hspace{-20mm} + a_{3} \, (w X \bar{w}) ( u w) (\bar{v} \bar{w})(u X \bar{v})^{2k-1} + a_{4} \, (w X \bar{v})(u X \bar{w})(uw)(\bar{v} \bar{w})(u X \bar{v})^{2k-2}  \nonumber \\ 	
		& \hspace{-20mm} + a_{5} \, (w X \bar{v})^{2} ( u X \bar{w}) (\bar{v} \bar{w})(u X \bar{v})^{2k-2} + a_{6} \, (u X \bar{w}) (w X \bar{w}) (w X \bar{v}) (u X \bar{v})^{2k-1}  \, ,
	\end{align}
	\item[] $\cZ_{2}$ structures:
	\begin{align}
		\cF_{2}(X; u, \bar{v}, w, \bar{w}) &= a_{7} \, (w X \bar{w})(uw)(\bar{v} \bar{w})(u X \bar{v})^{2k-2} + a_{8} \, (w X \bar{w})^{2} (u X \bar{v})^{2k-1} \nonumber \\
		& \hspace{15mm} + a_{9} \, (u X \bar{w})(w X \bar{v}) (w X \bar{w})(u X \bar{v})^{2k-2}\, ,
	\end{align}
	\item[] $\cZ_{3}$ structures:
	\begin{align}
		\cF_{3}(X; u, \bar{v}, w, \bar{w}) &= a_{10} \, (w X \bar{v}) (uw) (\bar{v} \bar{w}) (u X \bar{v})^{2k-2} + a_{11} \, (w X \bar{w})(w X \bar{v})(u X \bar{v})^{2k-1} \nonumber \\
		& \hspace{15mm} + a_{12} \, (w X \bar{v})^{2} (u X \bar{w}) (u X \bar{v})^{2k-2} \, ,
	\end{align}
	\item[] $\cZ_{4}$ structures:
	\begin{align}
		\cF_{4}(X; u, \bar{v}, w, \bar{w}) &= a_{13} \, (u X \bar{w}) (uw) (\bar{v} \bar{w}) (u X \bar{v})^{2k-2} + a_{14} \, (u X \bar{w})^{2} (w X \bar{v}) (u X \bar{v})^{2k-1} \nonumber \\
		& \hspace{15mm} + a_{15} \, (w X \bar{w}) (u X \bar{w}) (u X \bar{v})^{2k-1} \, ,
	\end{align}
	\item[] $\cZ_{5}$ structures:
	\begin{align}
		\cF_{5}(X; u, \bar{v}, w, \bar{w}) &= a_{16} \, (w X \bar{v}) (u X \bar{w}) (u X \bar{v})^{2k-1} + a_{17} \, (w X \bar{w}) (u X \bar{v})^{2k} \nonumber \\
		& \hspace{15mm} + a_{18} \, (u w) (\bar{v} \bar{w}) (u X \bar{v})^{2k-1} \, ,
	\end{align}
\end{itemize}
\end{subequations}
There are also the additional ``higher spin" structures, which appear only for $k > 1$:
\begin{subequations}
\begin{itemize}
	\item[] $\cZ_{2}$ structures:
	\begin{align} \label{QbQT higher-spin structures}
		\tilde{\cF}_{2}(X; u, \bar{v}, w, \bar{w}) &= a_{19} \, (uw)^{2} (\bar{v} \bar{w})^{2} (u X \bar{v})^{2k-3} + a_{20} \, (w X \bar{v})^{2} (u X \bar{w})^{2} (u X \bar{v})^{2k-3} \nonumber \\
		& \hspace{15mm} + a_{21} \, (u X \bar{w}) (w X \bar{v}) (u w) (\bar{v} \bar{w}) (u X \bar{v})^{2k-3}\, ,
	\end{align}
\end{itemize}
\end{subequations}
Hence, we will need to treat the cases $k=1$ and $k>1$ separately. First we will consider $k=1$, which corresponds to a field with the same properties as the supersymmetry current, $Q_{\a(2), \ad}$.

\subsection{Analysis for \texorpdfstring{$k=1$}{k=1}}

In this subsection we will determine the constraints on the coefficients for general $k$. First we must determine any linear dependence relations between the various polynomial structures. Using Mathematica it may be shown that $\cF_{\a \ad}(X; u, \bar{v}, w, \bar{w}) = 0$ for the following relations between the coefficients:
\begin{subequations}
	\begin{align}
		a_{3} &= a_{1} + a_{2} - a_{10} + a_{11}  \, , \\
		a_{5} &= - a_{1} - a_{4} + a_{10} + a_{12}  \, , \\
		a_{6} &= a_{1} - a_{2} + a_{4} - a_{10} - a_{12}  \, , \\
		a_{8} &= a_{7} + a_{10} - a_{11}  \, , \\
		a_{9} &=  - a_{7} - a_{10} - a_{12} \, ,\\
		a_{14} &= a_{10} + a_{12} - a_{13} \, , \\
		a_{15} &= - a_{10} + a_{11} + a_{13} \, , \\
		a_{17} &= - a_{11} - a_{12} - a_{16} \, , \\
		a_{18} &= - a_{10} - a_{12} - a_{16} \, .
	\end{align}
\end{subequations}
Therefore a linearly independent basis may be obtained by neglecting the structures corresponding to the coefficients $a_{1}, a_{2}, a_{4}, a_{7}, a_{10}, a_{11}, a_{12}, a_{13}, a_{16}$. There are only nine structures remaining, corresponding to the coefficients $a_{3}, a_{5}, a_{6}, a_{8}, a_{9}, a_{14}, a_{15}, a_{17}, a_{18}$ respectively. Now that we have identified any possible linear dependence between the polynomial structures, we impose the differential constraints and point-switch identities using Mathematica. After imposing the conservation equations \eqref{QbQT differential constraint 1b} and \eqref{QbQT differential constraint 2b}, we obtain the following relations between the coefficients:
\begin{align} 
	a_{6} &= \frac{1}{3} (  a_{3} - 2 a_{5} + 4 a_{18}  )  \, , &  a_{8} &= \frac{1}{3} ( - 2 a_{3} - 2 a_{5} + a_{18} ) \, , & a_{9} &= \frac{1}{3} ( 4 a_{3} + 4 a_{5} + 7 a_{18}) \, .
\end{align}
Hence, the differential constraints are sufficient to fix the correlation function up to four independent complex parameters, $a_{3}$, $a_{5}$, $a_{17}$ and $a_{18}$. The next constraint to impose is the reality condition \eqref{QbQT reality condition b}, from which we determine that the remaining four parameters must be purely real.

Finally we must check that the correlation function satisfies the differential constraint \eqref{QbQT differential constraint 3a} in accordance with conservation of the energy-momentum tensor. We begin by computing $\tilde{\cH}$ using \eqref{QbQT Htilde and Hc relation}; in the generating function formalism this may be written as
\begin{align}
	\tilde{\cH}(X; w, \bar{w}, v, \bar{v}, u, \bar{u}) = \frac{1}{(2k)!} X^{2k-3} (v X \pa \bar{s}) \, (\pa s X \bar{v})^{2k} \, \cH^{c}(X; u, \bar{u}, s, \bar{s}, w, \bar{w}) \, .
\end{align}
Conservation of the energy-momentum tensor at $x_{3}$ \eqref{QbQT differential constraint 3a} is now equivalent to imposing the following differential constraint on the tensor $\tilde{\cH}$:
\begin{align} \label{QbQT differential constraint 3b}
	\frac{\pa}{\pa X_{\sigma \dot{\sigma}}} \frac{\pa}{\pa w^{\sigma}} \frac{\pa}{\pa \bar{w}^{\dot{\sigma}}} \, \tilde{\cH}(X; w, \bar{w}, v, \bar{v}, u, \bar{u} ) = 0 \, . 
\end{align}
At this point we set $k=1$ and proceed with the analysis. Using Mathematica it may be shown that this constraint is automatically satisfied for the coefficient constraints above, hence, the correlation function $ \langle \bar{Q} Q T \rangle$ is determined up to four independent real parameters.

\subsection{Analysis for general \texorpdfstring{$k$}{k}}

Now let us carry out the analysis for general $k$; we must determine any linear dependence relations between the various polynomial structures. Indeed, we find that introducing the higher-spin contributions \eqref{QbQT higher-spin structures} results in the following supplementary linear dependence relation for $k > 1$, i.e. $\cF_{\a \ad}(X, u, \bar{v}, w, \bar{w}) = 0$ for the coefficient relations
\begin{align}
	a_{8} &= - a_{19} \, , & a_{9} &= a_{19} - a_{20} \, , & a_{21} &= - a_{19} - a_{20} \, .
\end{align}
Therefore the complete list of independent structures corresponds to the coefficients $a_{3}, a_{5}, a_{6}, a_{8}, a_{9}, a_{14}, a_{15}, a_{17}, a_{18}, a_{21}$. We now impose the differential constraints and point switch identities using Mathematica. After imposing the differential constraints arising from requiring conservation on the first and second point, that is \eqref{QbQT differential constraint 1b} and \eqref{QbQT differential constraint 2b} we obtain the $k$-dependent relations
\begin{subequations}
	\begin{align}
		a_{6} &= \frac{ a_{3} (-1 + k + 4 k^{2} - 4 k^{3} ) - 2 a_{5} (1 - k - 12 k^{2} + 12 k^{3} ) + a_{21} ( - 2 k + 8 k^{3}) }{ 1 + 3k - 16 k^{2} + 12 k^{3} } \, , \\[2mm]
		a_{8} &= \frac{  6 (1 - k)( 2 a_{3} k ( -1 + 2 k) + a_{5} (1 + 2k)) + a_{21} (-1 + 4k^{2}) }{2 + 6k - 32 k^{2} + 24 k^{3} } \, , \\[2mm]
		a_{9} &= \frac{ 5 a_{5} (- 1 + k) + a_{21} (1 - 4k)}{ 2 - 2 k } \, , \\[2mm]
		a_{18} &= \frac{ 2 (1 - k) ( 4 a_{3} k ( -1 + 2k) + a_{5} (1 + 12 k^{2})) }{ 2 + 6 k - 32 k^{2} + 24 k^{3} } \, .
	\end{align}
\end{subequations}
The remaining free coefficients are $a_{3}, a_{5}, a_{17}$ and $a_{21}$; the relations are also defined only for $k > 1$. Next we must impose the reality condition \eqref{QbQT reality condition b}, from which we find that the remaining coefficients must be purely real. Hence, we find that the correlation function is determined up to four independent real parameters. 

Finally, we must impose the differential constraint on $x_{3}$ which arises due to conservation of the energy-momentum tensor, that is, \eqref{QbQT differential constraint 3b}. 
Indeed, we have shown using Mathematica that \eqref{QbQT differential constraint 3b} is satisfied up to $k=4$, for higher values of $k$ the computations of 
$\tilde{\cH}$ seem to be beyond our computer power. 
However, we believe that the results will hold for higher values of $k$, so we can be reasonably confident that the correlation function is determined up to four independent real parameters for general $k$.


\section{Correlator \texorpdfstring{$\langle Q_{\a(2k) \ad }(x_{1}) \, Q_{\b(2k) \bd}(x_{2}) \, T_{\g(2) \gd(2)}(x_{3}) \rangle$}{< Q Q T >}}\label{section6}

In this section we will compute the correlation function $\langle Q Q T \rangle$. The ansatz for this correlator consistent with the general results of subsection \ref{subsection2.3} is
\begin{align} \label{QQT ansatz}
\langle Q_{\a(2k) \ad}(x_{1}) \, Q_{\b(2k) \bd}(x_{2}) \, T_{\g(2) \gd(2)}(x_{3}) \rangle 
&= \frac{1}{(x_{13}^{2} x_{23}^{2})^{k + \frac{5}{2}}} \, \cI_{\a(2k)}{}^{\ad'(2k)}(x_{13}) \, \bar{\cI}_{\ad}{}^{\a'}(x_{13}) \nonumber \\[-2mm] & \hspace{25mm} \times \cI_{\b(2k)}{}^{\bd'(2k)}(x_{23}) \, \bar{\cI}_{\bd}{}^{\b'}(x_{23}) \nonumber \\ & \hspace{25mm} \times \cH_{ \a' \ad'(2k) , \b' \bd'(2k), \g(2) \gd(2) }(X_{12}) \, , 
\end{align}
where $\cH$ is a homogeneous tensor field of degree $q = 4 - 2(k+\tfrac{5}{2}) = - 2k - 1 $. It is constrained as follows:
\begin{enumerate}
	\item[\textbf{(i)}] Under scale transformations of spacetime $x^{m} \mapsto x'^{m} = \l^{-2} x^{m}$ the three-point building blocks transform as $X^{m} \mapsto X'^{m} = \l^{2} X^{m}$. As a consequence, the correlation function transforms as 
	\begin{align}
	\langle Q_{\a(2k) \ad}(x'_{1}) \, Q_{\b(2k) \bd}(x'_{2}) \, T_{\g(2) \gd(2)}(x'_{3}) \rangle &= \nonumber \\ & \hspace{-30mm} (\l^{2})^{2k+9} \langle Q_{\a(2k) \ad}(x_{1}) \, Q_{\b(2k) \bd}(x_{2}) \, T_{\g(2) \gd(2)}(x_{3}) \rangle \, ,
	\end{align}
	which implies that $\cH$ obeys the scaling property
	\begin{equation}
	\cH_{ \a \ad(2k) , \b \bd(2k), \g(2) \gd(2) }(\l^{2} X) = (\l^{2})^{q} \, \cH_{ \a \ad(2k) , \b \bd(2k), \g(2) \gd(2) }(X) \, , \hspace{5mm} \forall \l \in \mathbb{R} \, \backslash \, \{ 0 \} \, .
	\end{equation}
	This guarantees that the correlation function transforms correctly under conformal transformations.
	
	\item[\textbf{(ii)}] The conservation of the fields $Q$ at $x_{1}$ and $x_{2}$ imply the following constraints on the correlation function:
	\begin{subequations}
		\begin{align}
		\pa_{(1)}^{\ad \a} \langle Q_{\a(2k-1) \a \ad}(x_{1}) \, Q_{\b(2k) \bd}(x_{2}) \, T_{\g(2) \gd(2)}(x_{3}) \rangle = 0 \, , \\
		\pa_{(2)}^{\bd \b} \langle Q_{\a(2k) \ad}(x_{1}) \, Q_{\b(2k-1) \b \bd}(x_{2}) \, T_{\g(2) \gd(2)}(x_{3}) \rangle = 0 \, .
		\end{align}
	\end{subequations}
	Using identities \eqref{Three-point building blocks - differential identities 2}, \eqref{Three-point building blocks - differential identities 3}, we obtain the following differential constraints on $\cH$:
	\begin{subequations}
		\begin{align}
		\pa_{X}^{\ad \a} \cH_{ \a \ad \ad(2k-1) , \b \bd(2k), \g(2) \gd(2) }(X) = 0 \, , \label{QQT differential constraint 1a}\\ 
		\pa_{X}^{\bd \b} \cH_{ \a \ad(2k) , \b \bd \bd(2k-1), \g(2) \gd(2) }(X) = 0 \, . \label{QQT differential constraint 2a}
		\end{align}
	\end{subequations}
	There is also a third constraint equation arising from conservation of $V$ at $x_{3}$,
	\begin{equation}
	\pa_{(3)}^{\gd \g} \langle Q_{\a(2k) \ad}(x_{1}) \, Q_{\b(2k) \bd}(x_{2}) \, T_{\g(2) \gd(2)}(x_{3}) \rangle = 0 \, .
	\end{equation}
	Similar to the previous example, we use the procedure outlined in subsection \ref{subsubsection2.3.1} and find the following relation between $\cH^{c}$ and $\tilde{\cH}$:
	\begin{align} \label{QQT Htilde and Hc relation}
	\tilde{\cH}_{ \g(2) \gd(2) , \b \bd(2k), \a(2k) \ad }(X) = X^{2k-3} \cI_{\b}{}^{\bd'}(X) \, \bar{\cI}_{\bd(2k)}{}^{\b'(2k)}(X) \, \cH^{c}_{ \a(2k) \ad , \b'(2k) \bd', \g(2) \gd(2) }(X) \, . 
	\end{align}
	Conservation on the third point is now equivalent to the following constraint on $\tilde{\cH}$:
	\begin{align} \label{QQT differential constraint 3a}
	\pa_{X}^{\dot{\s} \s} \tilde{\cH}_{ \s \dot{\s} \g \gd , \b \bd(2k), \a(2k) \ad }(X) = 0 \, . 
	\end{align}

	\item[\textbf{(iii)}] The correlation function possesses the following symmetry property under exchange of the fields at $x_{1}$ and $x_{2}$:
	\begin{equation}
	\langle Q_{\a(2k) \ad}(x_{1}) \, Q_{\b(2k) \bd}(x_{2}) \, T_{\g(2) \gd(2)}(x_{3}) \rangle = - \langle Q_{\b(2k) \bd}(x_{2}) \, Q_{\a(2k) \ad}(x_{1}) \, T_{\g(2) \gd(2)}(x_{3}) \rangle \, .
	\end{equation}
	This implies the following constraint on the tensor $\cH$:
	\begin{equation} \label{QQT point switch a}
	\cH_{ \a \ad(2k) , \b \bd(2k), \g(2) \gd(2) }(X) = - \cH_{ \b \bd(2k), \a \ad(2k), \g(2) \gd(2) }(-X)  \, .
	\end{equation}
\end{enumerate}

Hence, we have to solve for the tensor $\cH$ subject to the above constraints. Let us now streamline the calculations by constructing the generating function:
\begin{equation}
	\cH(X; u, \bar{u}, v, \bar{v}, w, \bar{w} ) = \cH_{ \a \ad(2k) , \b \bd(2k), \g(2) \gd(2) }(X) \, \mathbf{U}^{\a \ad(2k)} \mathbf{V}^{\b \bd(2k)} \mathbf{W}^{\g(2) \gd(2)} \, .
\end{equation}
%
The tensor $\cH$ is then obtained from the generating polynomial by acting on it with partial derivatives as follows:
\begin{equation}
	\cH_{ \a \ad(2k) , \b \bd(2k), \g(2) \gd(2) }(X)  = \frac{\pa}{ \pa \mathbf{U}^{\a \ad(2k)}} \frac{\pa}{ \pa \mathbf{V}^{\b \bd(2k)}} \frac{\pa}{ \pa  \mathbf{W}^{\g(2) \gd(2)}} \, \cH(X; u, \bar{u}, v, \bar{v}, w, \bar{w} )\, .
\end{equation}
%
Let us now convert our constraints on the tensor $\cH$ to constraints on the generating function. In particular, the differential constraints \eqref{QQT differential constraint 1a} and \eqref{QQT differential constraint 2a} become
\begin{subequations}
	\begin{align}
	\frac{\pa}{\pa X_{\sigma \dot{\sigma}}} \frac{\pa}{\pa u^{\sigma}} \frac{\pa}{\pa \bar{u}^{\dot{\sigma}}} \, \cH(X; u, \bar{u}, v, \bar{v}, w, \bar{w} ) &= 0 \, , \label{QQT differential constraint 1b} \\ 
	\frac{\pa}{\pa X_{\sigma \dot{\sigma}}} \frac{\pa}{\pa v^{\sigma}} \frac{\pa}{\pa \bar{v}^{\dot{\sigma}}} \, \cH(X; u, \bar{u}, v, \bar{v}, w, \bar{w} ) &= 0 \, , \label{QQT differential constraint 2b}
	\end{align}
\end{subequations}
while the homogeneity and point-switch constraints become:
\begin{subequations}
	\begin{align}
	\cH(\l^{2} X; u, \bar{u}, v, \bar{v}, w, \bar{w} ) &= (\l^{2})^{q} \, \cH(X; u, \bar{u}, v, \bar{v}, w, \bar{w} ) \, , \\
	\cH(X; u, \bar{u}, v, \bar{v}, w, \bar{w} ) &= - \cH(- X; v, \bar{v}, u, \bar{u}, w, \bar{w} ) \, . \label{QQT point switch b}
	\end{align}
\end{subequations}
Our task is now to construct the general solution for the polynomial $\cH$ consistent with the above constraints. The general expansion for $\cH$ is formed out of products of the basis objects \eqref{Basis scalar structures}. Let us start by decomposing the polynomial $\cH$, we have:
\begin{equation}
\cH(X; u, \bar{u}, v, \bar{v}, w, \bar{w} ) = \frac{1}{X^{2k+1}} \, u^{\a} v^{\b} \cF_{\a \b}(X; \bar{u}, \bar{v}, w, \bar{w} ) \, ,
\end{equation}
where have used the fact that $\cH$ is homogeneous degree 1 in both $u$ and $v$. The tensor $\cF$ is now homogeneous degree 0 in $X$, homogeneous degree 2 in $w$, $\bar{w}$, and homogeneous degree $2k$ in $\bar{u}$, $\bar{v}$. It may be decomposed into symmetric and anti-symmetric parts as follows:
\begin{equation}
	\cF_{\a \b}(X; \bar{u}, \bar{v}, w, \bar{w}) = \ve_{\a \b} A(X; \bar{u}, \bar{v}, w,  \bar{w}) + B_{(\a \b)}(X; \bar{u}, \bar{v}, w, \bar{w}) \, .
\end{equation}
It is straightforward to identify the possible structures in the expansion for $A$. We find
\begin{align}
	A(X; \bar{u}, \bar{v}, w,  \bar{w}) &= a_{1} \, (wX\bar{w})^{2} (\bar{u} \bar{v})^{2k} + a_{2} \, (w X \bar{u})^{2} (\bar{v} \bar{w})^{2} (\bar{u} \bar{v})^{2k-2} \nonumber \\
	& + a_{3} \, (w X \bar{w}) (w X \bar{u} ) (\bar{v} \bar{w} ) (\bar{u} \bar{v})^{2k-1} + a_{4} \, (w X \bar{w}) (w X \bar{v} ) (\bar{u} \bar{w} ) (\bar{u} \bar{v})^{2k-1} \nonumber \\
	& + a_{5} \, (w X \bar{v})^{2} (\bar{u} \bar{w} )^{2} (\bar{u} \bar{v})^{2k-2} + a_{6} \, (w X \bar{u}) (w X \bar{v} ) (\bar{u} \bar{w} ) (\bar{v} \bar{w} ) (\bar{u} \bar{v})^{2k-2} \, .
\end{align}
However, identifying all possible structures for the tensor $B$ is more challenging. To this end we introduce a basis of spinor structures, $Y_{i , \, \a}$:
\begin{align}
	Y_{1 , \, \a} &= w_{\a} \, , & Y_{2 , \, \a} &= \hat{X}_{\a \ad} \bar{u}^{\ad} \, , & Y_{3 , \, \a} &= \hat{X}_{\a \ad} \bar{v}^{\ad} \, , & Y_{4 , \, \a} &= \hat{X}_{\a \ad} \bar{w}^{\ad} \, .
\end{align}
From these basis spinors, we construct a set of symmetric objects, $\mathcal{Y}_{i j , \,  \a \b}$, defined as follows: 
\begin{align}
	\mathcal{Y}_{i j, \, \a \b} = \frac{1}{2} ( Y_{i , \, \a} Y_{j ,  \, \b} + Y_{i , \, \b} Y_{j , \, \a} ) \, .
\end{align}
These objects are symmetric in $\a, \b$, hence, they form a basis in which the tensor $B$ may be decomposed. However, since these objects are also symmetric in $i,j$, only 10 of them are unique, therefore we form the list $\cZ_{i, \, \a \b}$ out of the unique structures. We then have the decomposition
\begin{equation}
B_{(\a \b)}(X; \bar{u}, \bar{v}, w, \bar{w}) = \sum_{i=1}^{10} \cZ_{i , \, \a \b} \, B_{i}(X;  \bar{u}, \bar{v}, w, \bar{w}) \, ,
\end{equation}
where the polynomials $B_{i}$ are homogeneous degree 0 in $X$, with the appropriate homogeneity in $\bar{u}, \bar{v}, w, \bar{w}$. We now construct all possible polynomial structures for each $\cZ_{i , \, \a \b}$:
\vspace{-8mm}
\begin{subequations}
\begin{itemize}
	\item[] $\cZ_{1} := \mathcal{Y}_{11}$ structures:
	\begin{equation}
	B_{1}(X; \bar{u}, \bar{v}, w, \bar{w}) = b_{1} \, (\bar{u} \bar{w}) ( \bar{v} \bar{w}) (\bar{u} \bar{v})^{2k-1} \, ,
	\end{equation}
	\item[] $\cZ_{2} := \mathcal{Y}_{12}$ structures:
	\begin{align}
	B_{2}(X; \bar{u}, \bar{v}, w, \bar{w}) &= b_{2} \, (w X \bar{w}) ( \bar{v} \bar{w}) (\bar{u} \bar{v})^{2k-1} + b_{3} \, (w X \bar{u}) ( \bar{v} \bar{w})^{2} (\bar{u} \bar{v})^{2k-2} \nonumber \\
	& \hspace{15mm} + b_{4} \, (w X \bar{v}) (\bar{u} \bar{w}) ( \bar{v} \bar{w}) (\bar{u} \bar{v})^{2k-2} \, ,
	\end{align}
	\item[] $\cZ_{3} := \mathcal{Y}_{13}$ structures:
	\begin{align}
	B_{3}(X; \bar{u}, \bar{v}, w, \bar{w}) &= b_{5} \, (w X \bar{w}) ( \bar{u} \bar{w}) (\bar{u} \bar{v})^{2k-1} + b_{6} \, (w X \bar{v}) ( \bar{u} \bar{w})^{2} (\bar{u} \bar{v})^{2k-2} \nonumber \\
	& \hspace{15mm} + b_{7} \, (w X \bar{u}) (\bar{u} \bar{w}) ( \bar{v} \bar{w}) (\bar{u} \bar{v})^{2k-2} \, ,
	\end{align}
	\item[] $\cZ_{4} := \mathcal{Y}_{14}$ structures:
	\begin{align}
	B_{4}(X; \bar{u}, \bar{v}, w, \bar{w}) &= b_{8} \, (w X \bar{w}) (\bar{u} \bar{v})^{2k} + b_{9} \, (w X \bar{u}) ( \bar{v} \bar{w}) (\bar{u} \bar{v})^{2k-1} \nonumber \\
	& \hspace{15mm} + b_{10} \, (w X \bar{v}) (\bar{u} \bar{w}) (\bar{u} \bar{v})^{2k-1} \, ,
	\end{align}
	\item[] $\cZ_{5} := \mathcal{Y}_{22}$ structures:
	\begin{equation}
	B_{5}(X; \bar{u}, \bar{v}, w, \bar{w}) = b_{11} \, (w X \bar{w}) ( w X \bar{v}) ( \bar{v} \bar{w}) (\bar{u} \bar{v})^{2k-2} \, ,
	\end{equation}
	\item[] $\cZ_{6} := \mathcal{Y}_{23}$ structures:
	\begin{align}
	B_{6}(X; \bar{u}, \bar{v}, w, \bar{w}) &= b_{12} \, (w X \bar{w})^{2} (\bar{u} \bar{v})^{2k-1} + b_{13} \, (w X \bar{w}) ( w X \bar{u}) (\bar{v} \bar{w}) (\bar{u} \bar{v})^{2k-2} \nonumber \\
	& \hspace{15mm} + b_{14} \, (w X \bar{w}) (w X \bar{v}) (\bar{u} \bar{w}) (\bar{u} \bar{v})^{2k-2} \, ,
	\end{align}
	\item[] $\cZ_{7} := \mathcal{Y}_{24}$ structures:
	\begin{align}
	B_{7}(X; \bar{u}, \bar{v}, w, \bar{w}) &= b_{15} \, (w X \bar{w}) (w X \bar{v}) (\bar{u} \bar{v})^{2k-1} + b_{16} \, (w X \bar{u}) ( w X \bar{v}) (\bar{v} \bar{w}) (\bar{u} \bar{v})^{2k-2} \nonumber \\
	& \hspace{15mm} + b_{17} \, (w X \bar{v})^{2} (\bar{u} \bar{w}) (\bar{u} \bar{v})^{2k-2} \, ,
	\end{align}
	\item[] $\cZ_{8} := \mathcal{Y}_{33}$ structures:
	\begin{equation}
	B_{8}(X; \bar{u}, \bar{v}, w, \bar{w}) = b_{18} \, (w X \bar{w}) ( w X \bar{u}) ( \bar{u} \bar{w}) (\bar{u} \bar{v})^{2k-2} \, ,
	\end{equation}
	\item[] $\cZ_{9} := \mathcal{Y}_{34}$ structures:
	\begin{align}
	B_{9}(X; \bar{u}, \bar{v}, w, \bar{w}) &= b_{19} \, (w X \bar{w}) (w X \bar{u}) (\bar{u} \bar{v})^{2k-1} + b_{20} \, (w X \bar{u})^{2} ( \bar{v} \bar{w}) (\bar{u} \bar{v})^{2k-2} \nonumber \\
	& \hspace{15mm} + b_{21} \, (w X \bar{u}) (w X \bar{v}) (\bar{u} \bar{w}) (\bar{u} \bar{v})^{2k-2} \, ,
	\end{align}
	\item[] $\cZ_{10} := \mathcal{Y}_{44}$ structures:
	\begin{equation}
	B_{10}(X; \bar{u}, \bar{v}, w, \bar{w}) = b_{22} \, (w X \bar{u}) ( w X \bar{v}) (\bar{u} \bar{v})^{2k-1} \, .
	\end{equation}
\end{itemize}
\end{subequations}
There are also additional structures that are defined only for $k > 1$. Such structures will be denoted by $\tilde{B}$.
\begin{subequations}
\begin{itemize}
	\item[] $\cZ_{5}$ structures:
	\begin{equation} \label{QQT higher-spin structures a}
	\tilde{B}_{5}(X; \bar{u}, \bar{v}, w, \bar{w}) = b_{23} \, (w X \bar{v})^{2} ( \bar{u} \bar{w}) ( \bar{v} \bar{w}) (\bar{u} \bar{v})^{2k-3} + b_{24} \, (w X \bar{u}) ( w X \bar{v}) ( \bar{v} \bar{w})^{2} (\bar{u} \bar{v})^{2k-3} \, ,
	\end{equation}
	\item[] $\cZ_{6}$ structures:
	\begin{align} \label{QQT higher-spin structures b}
	\tilde{B}_{6}(X; \bar{u}, \bar{v}, w, \bar{w}) &= b_{25} \, (w X \bar{u})^{2} (\bar{v} \bar{w})^{2} (\bar{u} \bar{v})^{2k-3} + b_{26} \, (w X \bar{v})^{2} (\bar{u} \bar{w})^{2} (\bar{u} \bar{v})^{2k-3} \nonumber \\
	& \hspace{15mm} + b_{27} \, (w X \bar{u}) (w X \bar{v}) (\bar{u} \bar{w}) (\bar{v} \bar{w}) (\bar{u} \bar{v})^{2k-3} \, ,
	\end{align}
	\item[] $\cZ_{8}$ structures:
	\begin{equation} \label{QQT higher-spin structures c}
	\tilde{B}_{8}(X; \bar{u}, \bar{v}, w, \bar{w}) = b_{28} \, (w X \bar{u})^{2} ( \bar{u} \bar{w}) ( \bar{v} \bar{w}) (\bar{u} \bar{v})^{2k-3} + b_{29} \, (w X \bar{u}) ( w X \bar{v}) ( \bar{u} \bar{w})^{2} (\bar{u} \bar{v})^{2k-3}
	\end{equation}
\end{itemize}
\end{subequations}
Therefore we must analyse the $k=1$ and $k>1$ cases separately. 

\subsection{Analysis for \texorpdfstring{$k=1$}{k=1}}

First we must determine any linear dependence relations between the various polynomial structures. In this case, since there are many structures, the linear dependence relations are rather complicated. For the $A$ structures, we find $A(X; \bar{u}, \bar{v}, w,  \bar{w}) = 0$ for the choice of coefficients
\begin{align}
	a_{4} &= - a_{1} + a_{2} \, , & a_{5} &= - a_{1} - a_{3} \, , & a_{6} &= 2 a_{1} - a_{2} + a_{3} \, .
\end{align}
Hence, the structures corresponding to $a_{1}, a_{2}$ and $a_{3}$ may be neglected, and we are left with only the structures with coefficients $a_{4}, a_{5}, a_{6}$.
Next we find linear dependence amongst the $B$ structures, we find $B_{(\a \b)}(X; \bar{u}, \bar{v}, w,  \bar{w}) = 0$ for the choice
\begin{subequations}
	\begin{align}
		b_{6} &= b_{1} - b_{2} - b_{4} - b_{5} + b_{11} + b_{12} + b_{14}  \, , \\
		b_{7} &= - b_{1} + b_{2} - b_{3} + b_{5} - b_{12} + b_{13} + b_{18} \, , \\
		b_{8} &= - b_{1} + b_{2} + b_{4} - b_{10} - b_{11} + b_{15} + b_{17}  \, , \\
		b_{9} &= - b_{1} + b_{3} + b_{4} - b_{10} - b_{11} + b_{12} - b_{13} + b_{15} + b_{17} + b_{19} - b_{20}  \, , \\
		b_{21} &= - b_{11} - b_{13} - b_{14} - b_{16} - b_{17} - b_{18} - b_{20} \, , \\ b_{22} &= b_{11} - b_{12} + b_{13} - b_{15} + b_{16} - b_{19} + b_{20} \, .
	\end{align}
\end{subequations}
Therefore a linearly independent basis may be constructed out of the structures corresponding to the coefficients $b_{6}, b_{7}, b_{8}, b_{9}, b_{21}$ and $b_{22}$. Overall there are nine independent structures to consider. We now impose the differential constraints and point switch identities using Mathematica. After imposing \eqref{QQT differential constraint 2a},\eqref{QQT differential constraint 2b} we obtain the following relations between the coefficients:
\begin{subequations}
	\begin{align}
	b_{6} &= \frac{1}{90} ( - 126 a_{4} - 114 a_{5} - 115 a_{6} ) \, , \\
	b_{7} &= \frac{1}{90} ( 114 a_{4} + 126 a_{5} + 115 a_{6} ) \, , \\
	b_{8} &= \frac{1}{45} ( 42 a_{4} - 12 a_{5} + 5 a_{6} ) \, ,\\  
	b_{9} &= \frac{1}{90} ( - 54 a_{4} - 66 a_{5} - 95 a_{6} - 4 b_{21} - 44 b_{22} ) \, , \\
	b_{21} &= \frac{2}{3} ( a_{4} - a_{5} ) \, ,\\  
	b_{22} &= \frac{1}{6} ( - 2 a_{4} + 2 a_{5} + 5 a_{6} ) \, . 
	\end{align}
\end{subequations}
Hence, the differential constraints fix the correlation function up to three parameters. Next we must impose the point switch identity \eqref{QQT point switch b}, from which we obtain $a_{5} = a_{4}$, hence, we are left with the free complex parameters $a_{4}$ and $a_{6}$.

We must now impose \eqref{QQT differential constraint 3a} in accordance with conservation of the energy momentum tensor. First we compute $\tilde{\cH}$ using \eqref{QQT Htilde and Hc relation}, which in the generating function formalism may be written as
\begin{align} \label{QQT Htilde and Hc relation b}
\tilde{\cH}(X; w, \bar{w}, v, \bar{v}, u, \bar{u}) = \frac{1}{(2k)!} \, X^{2k-3} (v X \pa \bar{s}) \, (\pa s X \bar{v})^{2k} \, \cH^{c}(X; u, \bar{u}, s, \bar{s}, w, \bar{w}) \, ,
\end{align}
while the differential constraint \eqref{QQT differential constraint 3a} is equivalent to
\begin{align}
\frac{\pa}{\pa X_{\sigma \dot{\sigma}}} \frac{\pa}{\pa w^{\sigma}} \frac{\pa}{\pa \bar{w}^{\dot{\sigma}}} \, \tilde{\cH}(X; w, \bar{w}, v, \bar{v}, u, \bar{u} ) = 0 \, . \label{QQT differential constraint 3b}
\end{align}
At this point we can freely set $k=1$ and check whether our solution is consistent with conservation at $x_{3}$. Using Mathematica, it may be shown that \eqref{QQT differential constraint 3b} is satisfied provided that $a_{6} = - \frac{12}{5} a_{4}$, hence, the correlation function $ \langle Q Q T \rangle$ is determined up to a single complex parameter.

\subsection{Analysis for general \texorpdfstring{$k$}{k}}

Now let us complete the analysis for $k>1$. Again we must find a linearly independent basis of polynomial structures. If we supplement the set of basis structures corresponding to $b_{6}, b_{7}, b_{8}, b_{9}, b_{21}$ and $b_{22}$ with the $\tilde{B}$ structures defined in \eqref{QQT higher-spin structures a}, \eqref{QQT higher-spin structures b}, \eqref{QQT higher-spin structures c}, then it may be shown that $B_{(\a \b)}(X; \bar{u}, \bar{v}, w,  \bar{w}) = 0$ for the choice
\begin{subequations}
	\begin{align}
		b_{6} &= b_{23} + b_{26} \, , \\
		b_{7} &= - b_{21} - b_{22} + b_{23} + b_{24} + b_{27} \, , \\
		b_{8} &= - b_{23} \, , \\
		b_{9} &= - b_{22} - b_{23} - b_{24} \, , \\
		b_{25} &= - b_{22} - b_{24} \, , \\ 
		b_{28} &= b_{21} + 2 b_{22} - b_{23} - b_{27} \, , \\
		b_{29} &= - b_{21} - b_{22} - b_{26} \, .
	\end{align}
\end{subequations}
Hence, there are ten independent structures to consider, corresponding to the coefficients $a_{4}, a_{5}, a_{6}, b_{6}, b_{7}, b_{8}, b_{9}, b_{25}, b_{28}$ and $b_{29}$.
We now impose the differential constraints and point-switch identities; after imposing \eqref{QQT differential constraint 2a}, \eqref{QQT differential constraint 2b} and \eqref{QQT point switch b} we obtain $a_{5} = a_{4}$, supplemented by the following $k$-dependent relations between the $b$ coefficients:
\begin{subequations}
	\begin{align}
	b_{6} &= \frac{ 
		2 a_{4} (24 + 45 k + k^2 + 14 k^3 - 4 k^4 - 8 k^5) + a_{6} (24 + 39 k + 26 k^2 - 12 k^3 - 8 k^4) }{(-5 + 2 k) (1 + 2 k) (-2 + k + 5 k^2 + 2 k^3)}  \, , \\
	b_{7} &= - \frac{2 ( a_{4} (30 + 58 k - 8 k^2 - 8 k^3) + 
		a_{6} (15 + 23 k + 5 k^2 - 6 k^3 + 12 k^4 + 8 k^5))}{(-5 + 2 k) (1 + 2 k) (-2 + k + 5 k^2 + 2 k^3)}  \, , \\
	b_{8} &= \frac{-2 a_{4} (19 + 25 k - 30 k^2 - 4 k^3 + 8 k^4) + 
		a_{6} (-19 - 17 k + 2 k^2 + 20 k^3 + 8 k^4)}{(-5 + 2 k) (1 + 2 k) (-2 + k + 5 k^2 + 2 k^3)}  \, , \\
	b_{9} &= \frac{ 4 a_{4}(-7 - 7 k + 20 k^2 + 12 k^3) + 2 a_{6} (-7 + k + 8 k^2 + 4 k^3) }{(-5 + 2 k) (1 + 2 k) (-2 + k + 5 k^2 + 2 k^3)} \, , \\
	b_{25} &= \frac{k (3 + 2 k) (2 a_{4} (-3 + k + 2 k^2) - a_{6} (1 + 2 k) )}{10 - 9 k - 23 k^2 + 4 k^4} \, , \\
	b_{28} &= \frac{2 (3 + 2 k) (a_{4} (2 + 2 k - 4 k^2) + a_{6} (1 - k + k^2 + 2 k^3))}{10 - 9 k - 23 k^2 + 4 k^4} \, , \\
	b_{29} &= \frac{k (3 + 2 k) ( 2 a_{4} (-3 + k + 2 k^2) - a_{6} (1 + 2 k) )}{10 - 9 k - 23 k^2 + 4 k^4} \, .
	\end{align}
\end{subequations}
Hence, after imposing conservation on the first and second point, we find there are two free complex coefficients remaining. The last constraint to impose is conservation on $x_{3}$, that is \eqref{QQT differential constraint 3b}. We cannot obtain a relation for arbitrary $k$, as from a computational standpoint one must fix $k$ in order to compute $\tilde{\cH}$ as in \eqref{QQT Htilde and Hc relation b}. However, we find that the correlation function is fixed up to a single parameter up to $k=4$, after which the computations become incredibly long and beyond our computer power. 
For $k=2$ we find $a_{6} = - \frac{20}{7} a_{4}$, for $k=3$, $a_{6} = - \frac{28}{9} a_{4}$, and $k=4$, $a_{6} = -\frac{36}{11} a_{4}$. We anticipate that similar results will hold for general $k$ as well.


\section{Discussion on supersymmetry}\label{Discussion}

In this section we will concentrate on the case $k=1$, which corresponds to a ``supersym\-metry-like" current $Q_{\a \b \ad} = (\s^{m})_{\a \ad} Q_{m , \b}$ of dimension--$\frac{7}{2}$ satisfying the conservation equation
\be 
\partial^{\a\ad} Q_{\a \b \ad}=0\,. 
\label{zh1}
\ee
However, our analysis in the previous sections did not assume supersymmetry. The question that naturally arises is whether the ``supersymmetry-like" current actually is the supersymmetry current. That is, whether a conformal field theory possessing a conserved fermionic current of spin--$\frac{3}{2}$ is superconformal. 

In any supersymmetric field theory the supersymmetry current is a component of the supercurrent $J_{\a \ad} (z)$, which also contains the energy-momentum tensor. 
As was explained in the introduction, this implies that the three-point functions $\langle Q Q T \rangle $ and $\langle \bar Q Q T \rangle$ must be contained in the 
three-point function of the supercurrent $\langle J J J \rangle $. It is known that the general form of $\langle J J J\rangle $
is fixed by superconformal symmetry up to two independent structures~\cite{Osborn:1998qu}. 
Hence, this implies that in any superconformal field theory, $\langle Q Q T \rangle$ and $\langle \bar Q Q T \rangle $
must also be fixed up to at most two independent structures. Moreover, the three-point function $\langle Q Q T \rangle $ must actually vanish. Indeed, in a supersymmetric 
theory $Q$ carries an $R$-symmetry charge and, hence, the entire correlator $\langle Q Q T \rangle $ carries an $R$-symmetry charge. 
However, by performing a simple change of variables in the path integral it then follows that $\langle Q Q T \rangle =0$.
In addition, our analysis in Section \ref{section6} showed that, in general, conformal symmetry fixes $\langle Q Q T \rangle $ up one overall parameter, which is inconsistent 
with supersymmetry. We also found in Section \ref{section5} that the three-point function $\langle \bar Q Q T \rangle$ is fixed up to four rather than two independent parameters,
which, in general, is also inconsistent with the general form of $\langle J J J \rangle $. 

Similarly, we can examine the three-point functions $\langle \bar Q Q V \rangle $ and $\langle  Q Q V \rangle$ studied in Sections \ref{section3} and \ref{section4} respectively.
In supersymmetric theories, the vector current $V_{m}$ belongs to the flavour current multiplet $L (z)$. Hence, the correlation functions $\langle \bar{Q} Q V \rangle $ and $\langle Q Q V \rangle$ are contained 
in the three-point function $\langle J J L \rangle$. It is known~\cite{Osborn:1998qu} that $\langle J J L \rangle$ is fixed by superconformal symmetry up to an overall real coefficient. Hence, $\langle \bar Q Q V \rangle$ must also be fixed up to an overall coefficient. 
As for $\langle Q Q V \rangle $, it must vanish just like $\langle Q Q T \rangle $. However, our analysis in Sections \ref{section3}, \ref{section4} showed that $\langle \bar Q Q V \rangle$
is fixed up to three independent coefficients while $\langle Q Q V \rangle $ vanishes. The result for $\langle \bar{Q} Q V \rangle $ is therefore inconsistent with the general form of $\langle J J L \rangle$.


\section*{Acknowledgements}
The authors would like to thank Sergei Kuzenko and Jessica Hutomo for their comments on the main results, and Daniel Hutchings, Nowar Koning, Michael Ponds, Emmanouil Raptakis and Kai Turner for valuable discussions. The work of E.I.B. is supported in part by the Australian Research Council, project No. DP200101944.
The work of B.S. is supported by the \textit{Bruce and Betty Green Postgraduate Research Scholarship} under the Australian Government Research Training Program.



\appendix

\section{4D conventions and notation}\label{AppA}

Our conventions closely follow that of \cite{Buchbinder:1998qv}. For the Minkowski metric $\eta_{mn}$ we use the ``mostly plus'' convention: $\eta_{mn} = \text{diag}(-1,1,1,1)$. Spinor indices on spin-tensors are raised and lowered with the $\text{SL}(2,\mathbb{C})$ invariant spinor metrics
\begin{align}
\ve_{\a \b} = 
\begingroup
\setlength\arraycolsep{4pt}
\begin{pmatrix}
\, 0 & -1 \, \\
\, 1 & 0 \,
\end{pmatrix}
\endgroup 
\, , & \hspace{10mm}
\ve^{\a \b} =
\begingroup
\setlength\arraycolsep{4pt}
\begin{pmatrix}
\, 0 & 1 \, \\
\, -1 & 0 \,
\end{pmatrix}
\endgroup 
\, , \hspace{10mm}
\ve_{\a \g} \, \ve^{\g \b} = \d_{\a}{}^{\b} \, , \\[4mm]
\ve_{\ad \bd} = 
\begingroup
\setlength\arraycolsep{4pt}
\begin{pmatrix}
\, 0 & -1 \, \\
\, 1 & 0 \,
\end{pmatrix}
\endgroup 
\, , & \hspace{10mm}
\ve^{\ad \bd} =
\begingroup
\setlength\arraycolsep{4pt}
\begin{pmatrix}
\, 0 & 1 \, \\
\, -1 & 0 \,
\end{pmatrix}
\endgroup 
\, , \hspace{10mm}
\ve_{\ad \gd} \, \ve^{\gd \bd} = \d_{\ad}{}^{\bd} \, .
\end{align}
Given the spinor fields $\f_{\a}$, $\bar{\f}_{\ad}$, the spinor indices $\a = 1, 2$, $\ad = \bar{1}, \bar{2}$ are raised and lowered according to the following rules:
\begin{align}
	\f_{\a} &= \ve_{\a \b} \, \f^{\b} \, , & \f^{\a} &= \ve^{\a \b} \, \f_{\b} \, , & \bar{\f}_{\ad} &= \ve_{\ad \bd} \, \bar{\f}^{\b} \, , & \bar{\f}^{\ad} &= \ve^{\ad \bd} \, \bar{\f}_{\bd} \, .
\end{align}
It is also useful to introduce the complex $2 \times 2$ $\s$-matrices, defined as follows:
\begin{align}
	\s_{0} &= 
	\begingroup
	\setlength\arraycolsep{4pt}
	\begin{pmatrix}
	\, 1 & 0 \, \\
	\, 0 & 1 \,
	\end{pmatrix}
	\endgroup 
	\, , & \hspace{5mm}
	\s_{1} &=
	\begingroup
	\setlength\arraycolsep{4pt}
	\begin{pmatrix}
	\, 0 & 1 \, \\
	\, 1 & 0 \,
	\end{pmatrix}
	\endgroup 
	\, , & \hspace{5mm}
	\s_{2} &=
	\begingroup
	\setlength\arraycolsep{4pt}
	\begin{pmatrix}
	\, 0 & -\text{i} \, \\
	\, \text{i} & 0 \,
	\end{pmatrix}
	\endgroup 
	\, , & \hspace{5mm}
	\s_{3} &=
	\begingroup
	\setlength\arraycolsep{4pt}
	\begin{pmatrix}
	\, 1 & 0 \, \\
	\, 0 & -1 \,
	\end{pmatrix}
	\endgroup 
	\, .
\end{align}
The $\s$-matrices span the Lie group $\text{SL}(2, \mathbb{C})$, the universal covering group of the Lorentz group $\text{SO}(3,1)$. Now let $\s_{m} = (\s_{0}, \vec{\s} )$, we denote the components of $\s_{m}$ as $(\s_{m})_{\a \ad}$, and define:
\begin{equation}
	(\tilde{\s}_{m})^{\ad \a} \equiv \ve^{\ad \bd} \ve^{\a \b} (\s_{m})_{\b \bd} \, .
\end{equation}	
It can be shown that the $\s$-matrices possess the following useful properties:
\begin{align}
	(\s_{m} \tilde{\s}_{n} + \s_{n} \tilde{\s}_{m}  )_{\a}{}^{\b} &= - 2 \eta_{m n} \, \d_{\a}^{\b} \, , \\
	(\tilde{\s}_{m } \s_{n} + \tilde{\s}_{n} \s_{m}  )^{\ad}{}_{\bd} &= - 2 \eta_{m n} \, \d^{\ad}_{\bd} \, , \\
	\text{Tr}(\s_{m} \tilde{\s}_{n} ) &= - 2 \eta_{m n} \, , \\
	(\s^{m})_{\a \ad} (\tilde{\s}_{m})^{\bd \b} &= - 2 \d_{\a}^{\ad} \d_{\ad}^{\bd} \, .
\end{align}
The $\s$-matrices are then used to convert spacetime indices into spinor ones and vice versa according to the following rules:
\begin{equation}
	X_{\a \ad} = (\s^{m})_{\a \ad} X_{m} \, , \hspace{10mm} X_{m} = - \frac{1}{2} (\tilde{\s}_{m})^{\ad \a} X_{\a \ad} \, .
\end{equation}
	

	


\printbibliography[heading=bibintoc,title={References}]



\end{document}